\documentclass[12pt]{article}

\usepackage{cite}
\usepackage{axodraw}

\makeatletter
\if@twoside \oddsidemargin -17pt \evensidemargin 00pt
\else \oddsidemargin 00pt \evensidemargin 00pt
\fi
\expandafter\ifx\csname mathrm\endcsname\relax\def\mathrm#1{{\rm #1}}\fi

\renewcommand{\theequation}{\thesection.\arabic{equation}}
\newcounter{saveeqn}

\@addtoreset{equation}{section}

\makeatother

\def\nl{\nonumber\\}

\def\beq{\begin{equation}}
\def\eeq{\end{equation}}
\def\beqar{\begin{eqnarray}}
\def\eeqar{\end{eqnarray}}
\def\barr#1{\begin{array}{#1}}
\def\earr{\end{array}}
\def\bfi{\begin{figure}}
\def\efi{\end{figure}}
\def\btab{\begin{table}}
\def\etab{\end{table}}
\def\bce{\begin{center}}
\def\ece{\end{center}}
\def\nn{\nonumber}
\def\disp{\displaystyle}
\def\text{\textstyle}

\def\veps{\varepsilon}

\def\refeq#1{\mbox{(\ref{#1})}}
\def\reffi#1{\mbox{Fig.~\ref{#1}}}
\def\reffis#1{\mbox{Figs.~\ref{#1}}}
\def\refta#1{\mbox{Table~\ref{#1}}}

\def\refse#1{\mbox{Sect.~\ref{#1}}}

\def\refapp#1{\mbox{App.~\ref{#1}}}

\def\citere#1{\mbox{Ref.~\cite{#1}}}
\def\citeres#1{\mbox{Refs.~\cite{#1}}}

\newcommand{\GeV}{\unskip\,\mathrm{GeV}}
\newcommand{\MeV}{\unskip\,\mathrm{MeV}}
\newcommand{\TeV}{\unskip\,\mathrm{TeV}}

\def\mathswitchr#1{\relax\ifmmode{\mathrm{#1}}\else$\mathrm{#1}$\fi}

\newcommand{\PW}{\mathswitchr W}
\newcommand{\PZ}{\mathswitchr Z}

\newcommand{\PH}{\mathswitchr H}
\newcommand{\Pe}{\mathswitchr e}

\newcommand{\Pd}{\mathswitchr d}
\newcommand{\Pf}{f}

\newcommand{\Pl}{l}
\newcommand{\Pu}{\mathswitchr u}
\newcommand{\Ps}{\mathswitchr s}
\newcommand{\Pb}{\mathswitchr b}
\newcommand{\Pc}{\mathswitchr c}
\newcommand{\Pt}{\mathswitchr t}
\newcommand{\Pp}{\mathswitchr p}

\newcommand{\Pep}{\mathswitchr {e^+}}
\newcommand{\Pem}{\mathswitchr {e^-}}

\newcommand{\PWp}{\mathswitchr {W^+}}

\def\mathswitch#1{\relax\ifmmode#1\else$#1$\fi}

\newcommand{\Mf}{\mathswitch {m_\Pf}}
\newcommand{\Ml}{\mathswitch {m_\Pl}}

\newcommand{\MW}{\mathswitch {M_\PW}}

\newcommand{\MZ}{\mathswitch {M_\PZ}}
\newcommand{\MH}{\mathswitch {M_\PH}}
\newcommand{\Me}{\mathswitch {m_\Pe}}

\newcommand{\Md}{\mathswitch {m_\Pd}}
\newcommand{\Mu}{\mathswitch {m_\Pu}}
\newcommand{\Ms}{\mathswitch {m_\Ps}}
\newcommand{\Mc}{\mathswitch {m_\Pc}}
\newcommand{\Mb}{\mathswitch {m_\Pb}}
\newcommand{\Mt}{\mathswitch {m_\Pt}}
\newcommand{\GW}{\mathswitch {\Gamma_\PW}}

\newcommand{\scrs}{\scriptscriptstyle}
\newcommand{\sw}{\mathswitch {s_{\scrs\PW}}}
\newcommand{\cw}{\mathswitch {c_{\scrs\PW}}}

\newcommand{\GF}{\mathswitch {G_\mu}}

\renewcommand{\O}{{\cal O}}

\newcommand{\ri}{{\mathrm{i}}}

\newcommand{\rd}{{\mathrm{d}}}

\newcommand{\soft}{{\mathrm{soft}}}

\newcommand{\M}{{\cal {M}}}

\newcommand{\born}{{\mathrm{Born}}}
\newcommand{\virt}{{\mathrm{virt}}}
\newcommand{\coll}{{\mathrm{coll}}}
\newcommand{\sub}{{\mathrm{sub}}}
\newcommand{\ini}{{\mathrm{ini}}}
\newcommand{\fin}{{\mathrm{fin}}}
\newcommand{\fact}{{\mathrm{fact}}}
\newcommand{\nonfact}{{\mathrm{nonfact}}}
\newcommand{\res}{{\mathrm{res}}}
\newcommand{\rec}{{\mathrm{rec}}}
\newcommand{\PA}{{\mathrm{PA}}}

\def\Li{\mathop{\mathrm{Li}_2}\nolimits}

\def\Re{\mathop{\mathrm{Re}}\nolimits}

\def\arc{\mathop{\mathrm{arc}}\nolimits}

\def\draftdate{\relax}
\def\mda{\relax}
\def\mua{\relax}
\def\mla{\relax}
\def\draft{
\def\thtystars{******************************}
\def\sixtystars{\thtystars\thtystars}
\typeout{}
\typeout{\sixtystars**}
\typeout{* Draft mode!
         For final version remove \protect\draft\space in source file *}
\typeout{\sixtystars**}
\typeout{}
\def\draftdate{\today}
\def\mua{\marginpar[\boldmath\hfil$\uparrow$]%
                   {\boldmath$\uparrow$\hfil}%
                    \typeout{marginpar: $\uparrow$}\ignorespaces}
\def\mda{\marginpar[\boldmath\hfil$\downarrow$]%
                   {\boldmath$\downarrow$\hfil}%
                    \typeout{marginpar: $\downarrow$}\ignorespaces}
\def\mla{\marginpar[\boldmath\hfil$\rightarrow$]%
                   {\boldmath$\leftarrow $\hfil}%
                    \typeout{marginpar: $\leftrightarrow$}\ignorespaces}
\def\Mua{\marginpar[\boldmath\hfil$\Uparrow$]%
                   {\boldmath$\Uparrow$\hfil}%
                    \typeout{marginpar: $\Uparrow$}\ignorespaces}
\def\Mda{\marginpar[\boldmath\hfil$\Downarrow$]%
                   {\boldmath$\Downarrow$\hfil}%
                    \typeout{marginpar: $\Downarrow$}\ignorespaces}
\def\Mla{\marginpar[\boldmath\hfil$\Rightarrow$]%
                   {\boldmath$\Leftarrow $\hfil}%
                    \typeout{marginpar: $\Leftrightarrow$}\ignorespaces}
\overfullrule 5pt
\oddsidemargin -15mm
\marginparwidth 29mm
}

\makeatletter

\def\eqnarray{\stepcounter{equation}\let\@currentlabel=\theequation
\global\@eqnswtrue
\global\@eqcnt\z@\tabskip\@centering\let\\=\@eqncr
$$\halign to \displaywidth\bgroup\hskip\@centering
  $\displaystyle\tabskip\z@{##}$\@eqnsel&\global\@eqcnt\@ne
  \hskip 2\arraycolsep \hfil${##}$\hfil
  &\global\@eqcnt\tw@ \hskip 2\arraycolsep $\displaystyle\tabskip\z@{##}$\hfil
   \tabskip\@centering&\llap{##}\tabskip\z@\cr}
\def\appendix{\par
 \setcounter{section}{0} \setcounter{subsection}{0}
 \def\thesection{\Alph{section}}}

\newcommand{\lsim}
{\;\raisebox{-.3em}{$\stackrel{\displaystyle <}{\sim}$}\;}
\newcommand{\gsim}
{\;\raisebox{-.3em}{$\stackrel{\displaystyle >}{\sim}$}\;}

\def\dsl{\mathpalette\make@slash}
\def\make@slash#1#2{\setbox\z@\hbox{$#1#2$}%
  \hbox to 0pt{\hss$#1/$\hss\kern-\wd0}\box0}

\makeatother

\begin{document}

\thispagestyle{empty}
\def\thefootnote{\fnsymbol{footnote}}
\setcounter{footnote}{1}
\null
\draftdate\hfill BI-TP 2000/04 \\
\strut\hfill DESY 01-121 \\
\strut\hfill Edinburgh 2001/11 \\
\strut\hfill hep-ph/0109062
\vskip 0cm
\vfill
\begin{center}
{\Large \boldmath{\bf
Electroweak radiative corrections to \\[.5em]
W-boson production at hadron colliders}
\par} \vskip 2.5em
{\large
{\sc Stefan Dittmaier}%
\footnote{Heisenberg fellow of the Deutsche Forschungsgemeinschaft DFG.}
\\[1ex]
{\normalsize \it 
Deutsches Elektronen-Synchrotron DESY, \\
D-22603 Hamburg, Germany
}\\[2ex]
{\sc Michael Kr\"amer%
}\\[1ex]
{\normalsize \it Department of Physics and Astronomy, 
University of Edinburgh,\\ Edinburgh EH9 3JZ, Scotland}\\[2ex]
}
\par \vskip 1em
\end{center}\par
\vskip .0cm \vfill {\bf Abstract:} \par 
The complete set of electroweak ${\cal O}(\alpha)$ corrections to the
Drell--Yan-like production of W~bosons is calculated and compared to
an approximation provided by the leading term of an expansion about
the W-resonance pole. All relevant formulae are listed explicitly, and
particular attention is paid to issues of gauge invariance and the
instability of the W~bosons. A detailed discussion of numerical
results underlines the phenomenological importance of the electroweak
corrections to W-boson production at the Tevatron and at the
LHC. While the pole expansion yields a good description of resonance
observables, it is not sufficient for the high-energy tail of
transverse-momentum distributions, relevant for new-physics searches.
\par
\vskip 1cm
\noindent
September 2001
\par
\null
\setcounter{page}{0}
\clearpage
\def\thefootnote{\arabic{footnote}}
\setcounter{footnote}{0}

\section{Introduction}

The Drell--Yan-like production of W~bosons represents one of the
cleanest processes with a large cross section at the Tevatron and at
the LHC.  This reaction is not only well suited for a precise
determination of the W-boson mass $\MW$, it also yields valuable
information on the parton structure of the proton. Specifically, the
accuracy of $\approx$~$15$--$20\MeV$ \cite{ai99} in the $\MW$
measurement envisaged at the LHC will improve upon the precision of
$\approx$~$30\MeV$ to be achieved at LEP2
\cite{lep2repWmass} and Tevatron Run~II \cite{ai96}, and thus 
competes with the precision of the $\MW$ measurement expected at a
future $\Pep\Pem$ collider
\cite{Accomando:1998wt}. Concerning quark distributions, precise
measurements of rapidity distributions provide information over a wide
range in $x$ \cite{Martin:2000ww}; a measurement of the d/u ratio
would, in particular, be complementary to HERA results. The more
direct determination of parton--parton luminosities instead of single
parton distributions is even more precise \cite{Dittmar:1997md};
extracting the corresponding luminosities from Drell--Yan-like
processes allows one to predict related $q\bar q$ processes at the
per-cent level.

Owing to the high experimental precision outlined above, the
predictions for the processes $\Pp\Pp/{\mathswitchr
p}\bar{\mathswitchr p}\to\PW\to l\nu_l$ should match per-cent
accuracy;
for specific observables the required theoretical accuracy is even higher.
To this end, radiative corrections have to be included.  In
particular, it is important to treat final-state radiation carefully,
since photon emission from the final-state lepton significantly
changes the lepton momentum, which is used in the determination of the
W-boson mass.  A first step to include electroweak corrections was
already made in \citere{Berends:1985qa}, where effects of final-state
radiation in the W-boson decay stage were taken into account. Those
effects lead to a shift in the value of $\MW$ of the order of
50--150$\MeV$.  The approximation of \citere{Berends:1985qa} was
improved much later in \citere{Baur:1999kt}, where the 
electroweak ${\cal O}(\alpha)$ corrections to resonant W-boson
production \cite{Wackeroth:1997hz} were discussed for W~production at
the Tevatron in detail.  The ${\cal O}(\alpha)$ corrections are of the
order of the known next-to-next-to-leading order (NNLO) QCD
corrections
\cite{Hamberg:1991np} to Drell--Yan-like processes.  In ${\cal
O}(\alpha^2)$ only a study of two-photon radiation exists
\cite{Baur:2000hm}, while the virtual counterpart is completely
unknown.  Discussions of QCD corrections to Drell--Yan-like processes
can be found in
\citeres{Ellis:1997sc,Catani:2000zg} and references therein.

In this paper we present the complete calculation of the electroweak
${\cal O}(\alpha)$ corrections, including non-resonant contributions.
In particular, we compare the full ${\cal O}(\alpha)$ correction to a
{\it pole approximation} 
(similar to the one used in \citeres{Baur:1999kt,Wackeroth:1997hz})
that is based on the correction to the
production of resonant W~bosons. All relevant formulae are listed
explicitly. Moreover, a discussion of numerical results is presented
for the Tevatron (Run II) and for the LHC. Partial results of this 
analysis have already been presented in the LHC workshop report 
\cite{Haywood:1999qg}.

The paper is organized as follows.  In \refse{se:partoncs} we set our
conventions and provide analytical results for the parton-level
subprocess. In particular, we describe the calculation of the complete
and the ``pole-approximated'' ${\cal O}(\alpha)$
corrections. Different methods for treating the infrared and collinear
singularities are presented and compared with each other. The hadronic
cross section is discussed in \refse{se:ppcs}. In
Section~\ref{se:numres} we present numerical results for W-boson
production at the Tevatron and at the LHC. Our conclusions are given
in \refse{se:concl}. Finally, the appendices provide some
supplementary formulae.

\section{The parton process 
\boldmath{$u\bar d\to\PWp\to\nu_l l^+(+\gamma)$}}
\label{se:partoncs}

\subsection{Conventions and lowest-order cross section}
\label{se:born}

We consider the parton process
\beq
u(p_u,\tau_u) + \bar d(p_d,\tau_d) \;\to\;
\nu_l(k_n,-) + l^+(k_l,\tau_l) \;\; [+\gamma(k,\lambda)],
\eeq
where $u$ and $d$ generically denote the light up- and down-type quarks,
$u=\Pu,\Pc$ and $d=\Pd,\Ps$. The lepton $l$ represents $l=\Pe,\mu,\tau$.
The momenta and helicities of the corresponding particles are
given in brackets. 
The Mandelstam variables are defined by
\beq
\hat s = (p_u+p_d)^2, \quad
\hat t = (p_d-k_l)^2, \quad
\hat u = (p_u-k_l)^2, \quad
s_{\nu_l l} = (k_n+k_l)^2.
\eeq
We neglect the fermion masses $m_u$, $m_d$, $\Ml$ whenever possible,
i.e.\ we keep these masses only as regulators in the logarithmic mass 
singularities originating from collinear photon emission or exchange.
Obviously, we have $\hat s = s_{\nu_l l}$ for the non-radiative process
$u\bar d\to\nu_l l^+$.
In lowest order only the Feynman diagram shown in \reffi{fig:borndiag}
contributes to the scattering amplitude, 
\bfi
\centerline{\unitlength 1pt
\begin{picture}(120,100)(0,0)
\ArrowLine( 5, 95)(30,50)
\ArrowLine(30, 50)( 5, 5)
\Photon(30,50)(100,50){2}{6}
\Vertex(30,50){1.5}
\Vertex(100,50){1.5}
\ArrowLine(100,50)(125, 95)
\ArrowLine(125, 5)(100, 50)
\put(58,32){$\PWp$}
\put(-8,90){u}
\put(-8, 0){$\bar d$}
\put(130,90){$\nu_l$}
\put(130, 0){$l^+$}
\end{picture}
}
\caption{Lowest-order diagram for $u\bar d\to\PWp\to\nu_l l^+$.}
\label{fig:borndiag}
\efi
and the Born amplitude reads
\beq
\M_0 = \frac{e^2 V^*_{ud}}{2\sw^2} \,
\left[ \bar v_d\gamma^\mu\omega_-u_u\right] \,
\disp\frac{1}{\hat s-\MW^2+\ri\MW\GW(\hat s)} \,
\left[ \bar u_{\nu_l}\gamma_\mu\omega_-v_l\right],
\label{eq:m0}
\eeq
with an obvious notation for the Dirac spinors $\bar v_d$, etc., and
the left-handed chirality projector $\omega_-=\frac{1}{2}(1-\gamma_5)$.
The electric unit charge is denoted by $e$, the weak mixing angle is 
fixed by the ratio $\cw^2=1-\sw^2=\MW^2/\MZ^2$ of the W- and Z-boson
masses $\MW$ and $\MZ$, and $V_{ud}$ is the CKM matrix element for
the $ud$ transition.

Strictly speaking, Eq.~\refeq{eq:m0} already goes beyond lowest order,
since the W-boson width $\GW(\hat s)$ results from the Dyson summation of
all insertions of the (imaginary part of the) W~self-energy. Defining
the mass $\MW$ and the width $\GW$ of the W~boson in the on-shell scheme
(see e.g.\ \citere{Denner:1993kt}), the Dyson summation directly leads to a 
{\it running width}, i.e.\
\beq
\GW(\hat s)\big|_{\mathrm{run}} = \GW \frac{\hat s}{\MW^2}.
\eeq
On the other hand, a description of the resonance by an expansion about
the complex pole in the complex $\hat s$ plane corresponds to a
{\it constant width}, i.e.\
\beq
\GW(\hat s)\big|_{\mathrm{const}} = \GW.
\eeq
In lowest order these two parametrizations of the resonance region are
fully equivalent, but the corresponding values of the line-shape
parameters $\MW$ and $\GW$ differ in higher orders 
\cite{Bardin:1988xt,Wackeroth:1997hz,Beenakker:1997kn},
\beq
\MW^2\big|_{\mathrm{const}} = 
\left.\left(\MW^2-\GW^2+\dots\right)\right|_{\mathrm{run}}, 
\qquad
\GW\big|_{\mathrm{const}} =
\GW\biggl(1-\frac{\GW^2}{2\MW^2}+\dots\biggr)\bigg|_{\mathrm{run}}. 
\label{eq:mwdef}
\eeq
Since $\MW|_{\mathrm{run}}-\MW|_{\mathrm{const}}\approx 26\MeV$, it is
necessary to state explicitly which parametrization is used in a
precision determination of the W-boson mass from the W~line-shape.

The differential lowest-order cross section is easily obtained by
squaring the lowest-order matrix element $\M_0$ of \refeq{eq:m0},
\beq
\biggl(\frac{\rd\hat\sigma_0}{\rd\hat\Omega}\biggr) 
= \frac{1}{12} \, \frac{1}{64\pi^2\hat s} \, |\M_0|^2
= \frac{\alpha^2 |V_{ud}|^2}{48\sw^4 \hat s}
\frac{\hat u^2}{|\hat s-\MW^2+\ri\MW\GW(\hat s)|^2},
\eeq
where the explicit factor $1/12$ results from the average over the
quark spins and colours, and $\hat\Omega$ is the solid angle of the
outgoing $l^+$ in the parton centre-of-mass (CM) frame.  The
electromagnetic coupling $\alpha=e^2/(4\pi)$ can be set to different
values according to different input-parameter schemes.  It can be
directly identified with the fine-structure constant $\alpha(0)$ or
the running electromagnetic coupling $\alpha(Q^2)$ at a high-energy
scale $Q$.  For instance, it is possible to make use of the value of
$\alpha(\MZ^2)$ that is obtained by analyzing \cite{Burkhardt:1995tt}
the experimental ratio
$R=\sigma(\Pep\Pem\to\mbox{hadrons})/(\Pep\Pem\to\mu^+\mu^-)$.  These
choices are called {\it $\alpha(0)$-scheme} and {\it
$\alpha(\MZ^2)$-scheme}, respectively, in the following.  Another
value for $\alpha$ can be deduced from the Fermi constant $\GF$,
yielding $\alpha_{\GF}=\sqrt{2}\GF\MW^2\sw^2/\pi$; this choice is
referred to as {\it $\GF$-scheme}.  The differences between these
schemes will become apparent in the discussion of the corresponding
${\cal O}(\alpha)$ corrections.

\subsection{Virtual corrections}

The virtual one-loop corrections comprise contributions of the
transverse part of the $W$~self-energy $\Sigma^W_{\mathrm{T}}$, 
corrections to the two $Wdu$ and $W\nu_l l$ vertices, 
box diagrams, and counterterms. The explicit expression for 
$\Sigma^W_{\mathrm{T}}$ (in the `t~Hooft--Feynman gauge) can, e.g., 
be found in \citere{Denner:1993kt}. The diagrams for the vertex and box
corrections, which are shown in \reffi{fig:vertboxdiags}, were
calculated using standard methods. 
\bfi
\centerline{
\setlength{\unitlength}{1cm}
\begin{picture}(16,8.0)
\put(0,5.5){\includegraphics{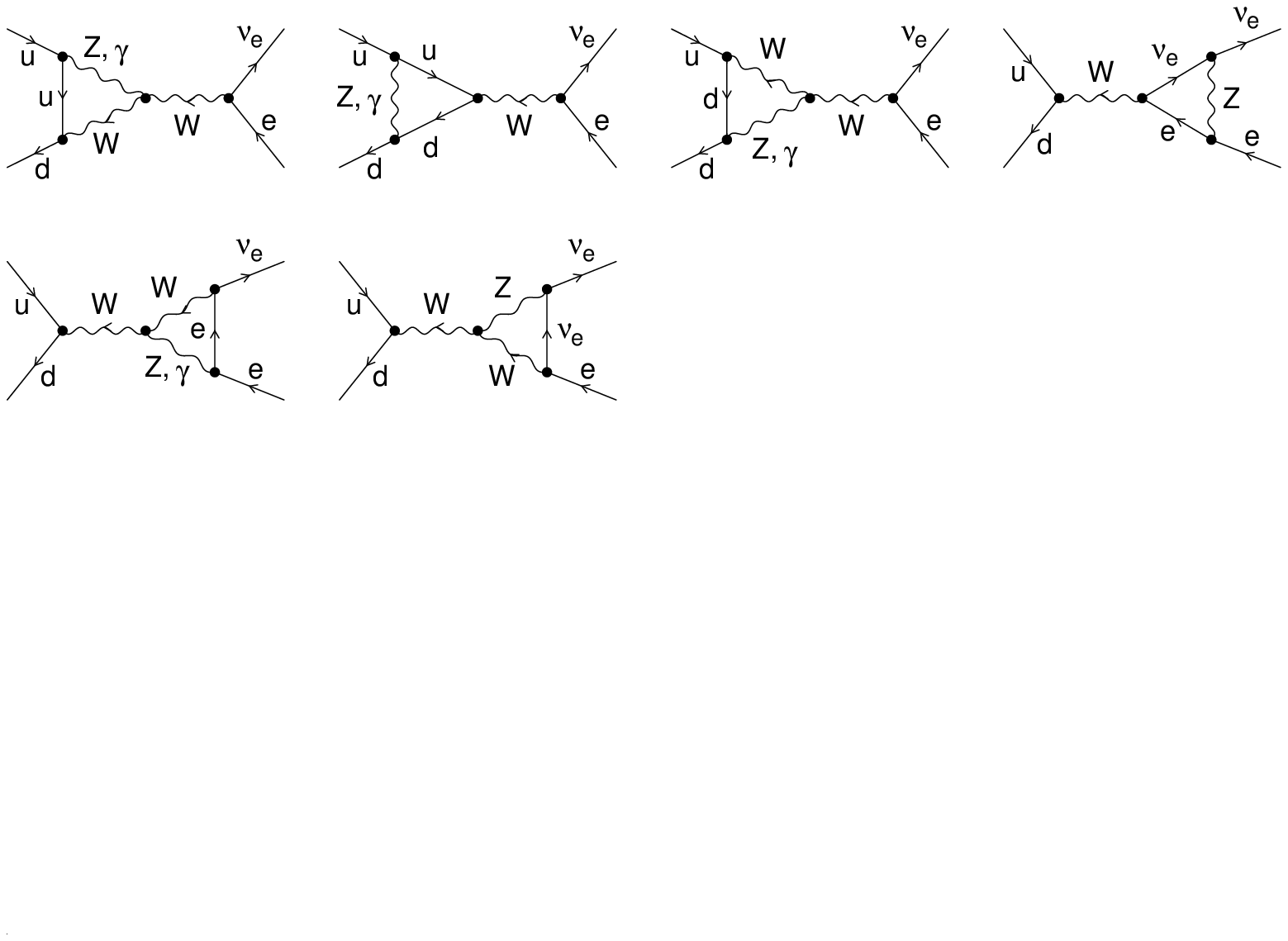}}
\put(0,0){\includegraphics{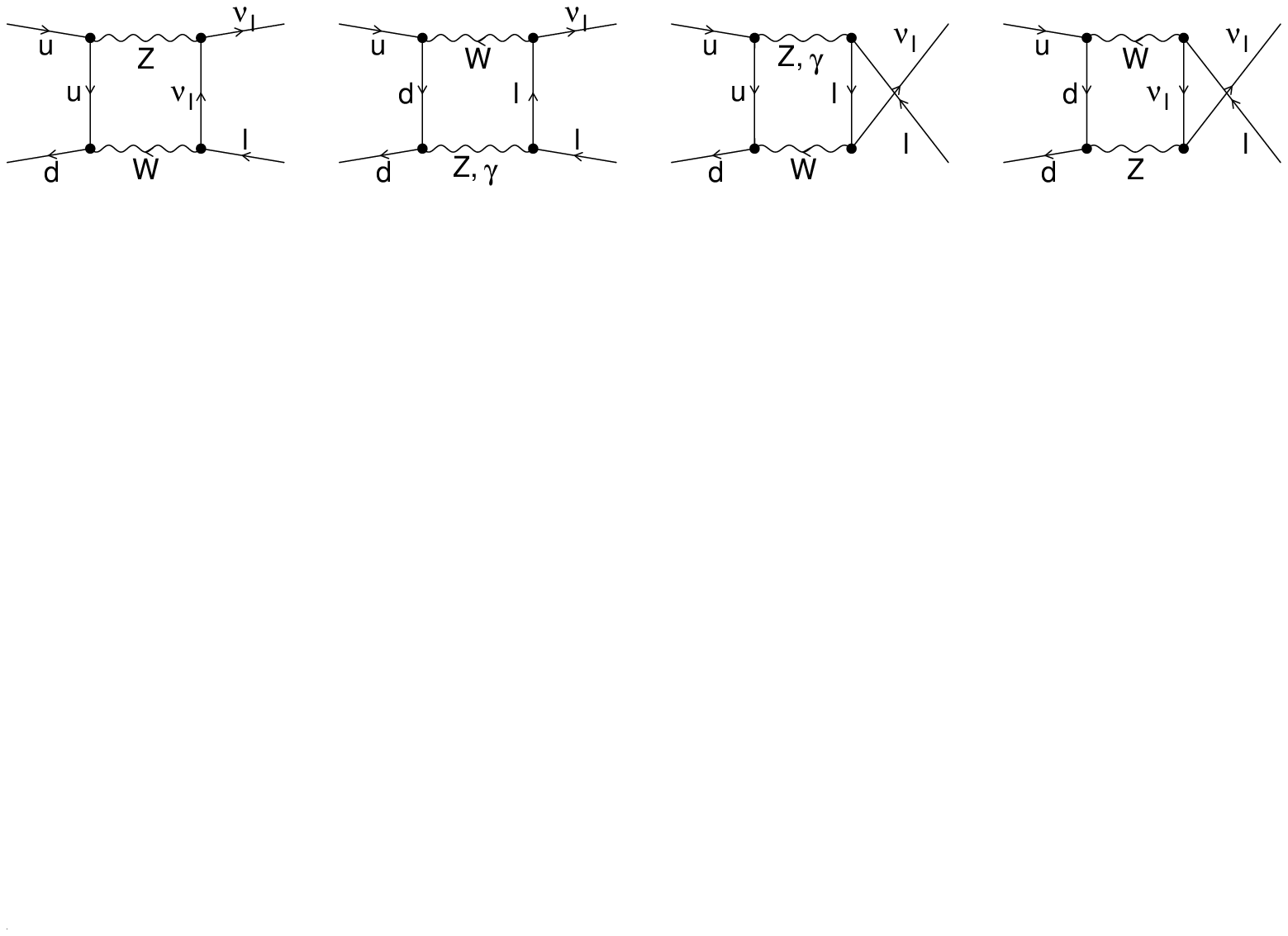}}
\end{picture} 
}
\caption{Diagrams for vertex and box corrections.}
\label{fig:vertboxdiags}
\efi
The Feynman diagrams and amplitudes were generated with {\sl FeynArts}
\cite{Kublbeck:1990xc}. The subsequent algebraic reduction
\cite{Passarino:1979jh} of the one-loop tensor integrals to scalar
integrals was performed with {\sl FeynCalc} \cite{Mertig:1991an}, and
the scalar integrals were evaluated using the methods and results of
\citere{'tHooft:1979xw}. The algebraic part was checked numerically 
by a completely independent calculation, in which the amplitudes are
expressed in terms of tensor coefficients using {\sl Mathematica}, and
the tensor reduction is done numerically. UV divergences are treated
in dimensional regularization, and the IR singularity is regularized
by an infinitesimal photon mass $m_\gamma$. The actual calculation is
performed in `t~Hooft--Feynman gauge using the on-shell
renormalization scheme described in
\citere{Denner:1993kt}, where, in particular, all renormalization
constants used in this paper can be found.  As an additional check we
have repeated the calculation within the background-field formalism
\cite{Denner:1995xt} and found perfect agreement.  In the following we
sketch the structure of the virtual corrections and emphasize those
points that are relevant for the treatment of the resonance and for
the change from one input-parameter scheme to another.  
The complete expressions
for the vertex and box corrections are provided in
\refapp{app:virtRCs}.

While the self-energy and vertex corrections are proportional to the
Dirac structure appearing in the lowest-order matrix element $\M_0$,
the calculation of the box diagrams leads to additional combinations of
Dirac chains. However, since the box diagrams are UV-finite, the
four-dimensionality of space-time can be used to reduce all Dirac
structures to the one of $\M_0$, see \refapp{app:box}. In summary, the
complete one-loop amplitude $\M_1$ can be expressed in terms of a
correction factor $\delta^{\virt}$ times the lowest-order matrix
element,
\beq
\M_1 = \delta^{\virt} \M_0.
\label{eq:m1}
\eeq
Thus, in ${\cal O}(\alpha)$ the squared matrix element reads
\beq
|\M_0+\M_1|^2 = (1+2\Re\{\delta^{\virt}\}) |\M_0|^2 + \dots,
\label{eq:dm2virt}
\eeq
so that the Breit--Wigner factors are completely contained in the
lowest-order factor $|\M_0|^2$.
Note that the Dyson-summed imaginary
part of the W~self-energy, which appears as $\GW(\hat s)$ in $\M_0$,
is not double-counted, since only the real part of
$\Sigma^W_{\mathrm{T}}$ enters $\Re\{\delta^{\virt}\}$ in ${\cal
O}(\alpha)$.%
\footnote{Note also that only the imaginary parts
induced by the fermion loops, which lead to the physical decay width in 
the resonance propagator, are resummed, while all other parts of 
$\Sigma^W_{\mathrm{T}}$ are treated in ${\cal O}(\alpha)$. 
In this way the differences between the on-shell renormalization
of the W-boson mass (see e.g.\ \citere{Denner:1993kt})
and other variants based on the complex pole
position in the inverse W~propagator are of ${\cal O}(\alpha^2)$, 
and thus beyond the accuracy of our calculation. In particular,
the problems \cite {Passera:1998uj} with the resummation of loops 
containing photon exchange are avoided.}
The correction factor $\delta^{\virt}$ is decomposed
into four different parts,
\beq
\delta^{\virt} = \delta_{WW}(\hat s) + 
\delta_{Wdu}(\hat s) + \delta_{W\nu_l l}(\hat s) + 
\delta_{\mathrm{box}}(\hat s,\hat t),
\label{eq:dvirt}
\eeq
according to the splitting into self-energy, vertex, and box diagrams.

The $W$~self-energy correction reads
\beq
\delta_{WW}(\hat s) = 
-\frac{\Sigma^W_{\mathrm{T}}(\hat s)-\delta\MW^2}{\hat s-\MW^2}
-\delta Z_W,
\eeq
where the explicit expression for the unrenormalized self-energy
$\Sigma^W_{\mathrm{T}}$ is given in Eq.~(B.4) of \citere{Denner:1993kt}.
In the on-shell renormalization scheme the renormalization constants 
for the W-boson mass and field, $\delta\MW^2$ and $\delta Z_W$, are 
directly related to $\Sigma^W_{\mathrm{T}}$.

The vertex corrections are given by
\beq
\delta_{Wdu}(\hat s)  = 
F_{Wdu}(\hat s) + \delta_{Wdu}^{\mathrm{ct}},
\qquad
\delta_{W\nu_l l}(\hat s) = 
F_{W\nu_l l}(\hat s) + \delta_{W\nu_l l}^{\mathrm{ct}},
\label{eq:dvert}
\eeq
where the explicit expression for the form factor $F_{Wff'}(\hat s)$ is
given in \refapp{app:vert}. The coun\-ter\-term $\delta_{Wff'}^{\mathrm{ct}}$
for the $Wff'$ vertex depends on the input-parameter scheme. In the
$\alpha(0)$-scheme (i.e.\ the usual on-shell scheme), it is given by%
\footnote{We consistently set the CKM matrix to the unit matrix in the
correction factor $\delta^{\virt}$, since mixing effects in the ${\cal
O}(\alpha)$ corrections are negligible. This means that the CKM matrix
appears only in the global factor $|V_{ud}|^2$ to the ${\cal
O}(\alpha)$-corrected parton cross section.}
\beq
\delta_{Wff'}^{\mathrm{ct}}\big|_{\alpha(0)} = 
\delta Z_e - \frac{\delta\sw}{\sw} + 
\frac{1}{2}(\delta Z_W + \delta Z^{f,\mathrm{L}} + \delta Z^{f',\mathrm{L}}).
\eeq
The wave-function renormalization constants $\delta Z^{f,\mathrm{L}}$
and $\delta Z^{f',\mathrm{L}}$ are obtained from the (left-handed part
of) the fermion self-energies, and the renormalization of the weak
mixing angle, i.e.\ $\delta\sw$, is connected to the mass
renormalization of the gauge-boson masses.  The charge renormalization
constant $\delta Z_e$ contains logarithms of the light-fermion masses,
inducing large corrections proportional to $\alpha\ln(m_f^2/\hat s)$,
which are related to the running of the electromagnetic coupling
$\alpha(Q^2)$ from $Q=0$ to a high-energy scale. In order to render
these quark-mass logarithms meaningful, it is necessary to adjust
these masses to the asymptotic tail of the hadronic contribution to
the vacuum polarization $\Pi^{AA}$ of the photon.  Using
$\alpha(\MZ^2)$, as defined in \citere{Burkhardt:1995tt}, as input
this adjustment is implicitly incorporated, and the counterterm reads
\beq
\delta_{Wff'}^{\mathrm{ct}}\big|_{\alpha(\MZ^2)} =
\delta_{Wff'}^{\mathrm{ct}}\big|_{\alpha(0)} -
\frac{1}{2}\Delta\alpha(\MZ^2),
\eeq
where 
\beq
\Delta\alpha(Q^2) = \Pi^{AA}_{f\ne \Pt}(0)-\Re\{\Pi^{AA}_{f\ne \Pt}(Q^2)\},
\eeq
\begin{sloppypar}
\noindent
with $\Pi^{AA}_{f\ne \Pt}$ denoting the photonic vacuum polarization
induced by all fermions other than the top quark (see also
\citere{Denner:1993kt}). In contrast to the $\alpha(0)$-scheme the 
coun\-ter\-term $\delta_{Wff'}^{\mathrm{ct}}\big|_{\alpha(\MZ^2)}$
does not involve light quark masses, since all corrections of the form
$\alpha^n\ln^n(m_f^2/\hat s)$ are absorbed in the lowest-order cross
section parametrized by
$\alpha(\MZ^2)=\alpha(0)/[1-\Delta\alpha(\MZ^2)]$.  In the
$\GF$-scheme, the transition from $\alpha(0)$ to $\GF$ is ruled by the
quantity $\Delta r$ \cite{Sirlin:1980nh,Denner:1993kt}, which is
deduced from muon decay,
\end{sloppypar}
\beq
\alpha_{\GF}=\frac{\sqrt{2}\GF\MW^2\sw^2}{\pi}
=\alpha(0)(1+\Delta r) \;+\; {\cal O}(\alpha^3).
\eeq
Therefore, the counterterm $\delta_{Wff'}^{\mathrm{ct}}$ reads
\beq
\delta_{Wff'}^{\mathrm{ct}}\big|_{\GF} =
\delta_{Wff'}^{\mathrm{ct}}\big|_{\alpha(0)} - \frac{1}{2}\Delta r.
\eeq
Since $\Delta\alpha(\MZ^2)$ is explicitly contained in $\Delta r$, the
large fermion-mass logarithms are also resummed in the $\GF$-scheme. Moreover,
the lowest-order cross section in $\GF$-parametrization absorbs large 
universal corrections induced by the $\rho$-parameter.

The box correction $\delta_{\mathrm{box}}(\hat s,\hat t)$ is the only
virtual correction that depends also on the scattering angle, i.e.\ on
the variables $\hat t$ and $\hat u=-\hat s-\hat t$. The explicit
expression for $\delta_{\mathrm{box}}(\hat s,\hat t)$ is given in
\refapp{app:box}.

Despite the separation of the resonance pole $(\hat s-\MW^2)^{-1}$ 
from the correction factor $\delta^{\virt}$ in \refeq{eq:m1},
$\delta^{\virt}$ still contains logarithms $\ln(\hat s-\MW^2+\ri\epsilon)$ 
that are singular on resonance. Since these singularities would be cured 
by a Dyson summation of the $W$~self-energy inside the loop diagrams, we
substitute
\beq
\ln(\hat s-\MW^2+\ri\epsilon) \;\to\; \ln(\hat s-\MW^2+\ri\MW\GW) 
\label{eq:onshelllogs}
\eeq
with a fixed width everywhere, independent of the use of a fixed or 
running width in lowest order. In principle, also a running width
could be used on the r.h.s., but the difference to the fixed width
is of two-loop order, and thus beyond the accuracy of our calculation.
The substitution \refeq{eq:onshelllogs}
does not disturb the gauge-invariance properties
of the one-loop amplitude $\M_1$, i.e.\ its gauge-parameter independence
and the validity of SU(2)$\times$U(1) Ward identities. The reason
is that the sum of all terms in $\M_1$ proportional to 
$\ln(\hat s-\MW^2+\ri\epsilon)$ separately fulfills the (algebraic)
relations arising from gauge invariance, since this logarithm
is the only resonant (and thus a unique) term in the relative correction 
$\delta^{\virt}$.

\subsection{Virtual correction in pole approximation}
 
If one is only interested in the production of (nearly) resonant
W~bosons, the electroweak corrections can be approximated by an
expansion \cite{Stuart:1991xk} about the resonance pole, which is
located in the complex $\hat s$ plane at $\MW^2-\ri\MW\GW$ up to
higher-order terms.  The approximation of taking into account only the
leading term of this expansion is called {\it pole approximation} (PA)
and should not be confused with the on-shell approximation for the
W~bosons.  In contrast to the PA, the on-shell approximation,
where the W~bosons are assumed to be stable, does not provide a
description of the W~line-shape. In the following we construct a PA
for the virtual correction factor $\delta^{\virt}$ in the same
way as a {\it double-pole approximation} was constructed in
\citere{Denner:2000bj} for the more complicated case of
W-pair production in $\Pep\Pem$ annihilation, $\Pep\Pem\to\PW\PW\to
4\,$fermions.
In this formulation the PA is only applied to the virtual corrections,
while the real corrections are based on the full photon-emission
matrix element, as described in the next section.
In principle, it is also possible to construct a PA for the real 
corrections, as for instance described in \citere{Jadach:1998hi}
for $\Pe\Pe\to\PW\PW$.%
\footnote{More details about the pole expansions that are used in
practice are reviewed in \citere{Grunewald:2000ju}.}
However, we prefer to
make use of the full matrix elements, as it was also done in the
variant of the PA used in \citere{Baur:1999kt} for W~production in
hadronic collisions.

In the PA the virtual corrections to $u\bar d\to\PW\to\nu_l l^+$ can
be classified into two categories. The first category comprises the
corrections to the production and the decay of an on-shell W~boson.
Owing to the independence of these subprocesses, these corrections are
called {\it factorizable}. All contributing Feynman graphs are of the
generic form shown in \reffi{fig:fact}.
\bfi
\centerline{
{\unitlength 1pt 
\begin{picture}(150,80)(0,0)
\ArrowLine( 5, 75)(25,40)
\ArrowLine(25, 40)( 5, 5)
\Photon(25,40)(125,40){2}{9}
\Vertex(25,40){1.5}
\Vertex(125,40){1.5}
\ArrowLine(125,40)(145, 75)
\ArrowLine(145, 5)(125, 40)
\GCirc(25,40){10}{0.5}
\GCirc(75,40){10}{1}
\GCirc(125,40){10}{0.5}
\DashLine(55,0)(55,80){2}
\DashLine(95,0)(95,80){2}
\put(-50,-10){\normalsize on-shell production}
\put(105,-10){\normalsize on-shell decay}
\end{picture}
} } 
\vspace*{.5em}
\caption{Generic diagram for factorizable corrections.}
\label{fig:fact}
\efi
By definition, the factorizable corrections receive only contributions
from the $W$~self-energy and the $Wff'$ vertex corrections. 
The corresponding correction factor $\delta^{\virt}_{\fact}$ is obtained
from $\delta_{WW}(\hat s)$ and $\delta_{Wff'}(\hat s)$ by setting
$\hat s=\MW^2$ and $\GW=0$. Since we have
$\delta_{WW}(\MW^2)|_{\GW=0}=0$ in the complete on-shell
renormalization scheme, we get
\beq
\delta^{\virt}_{\fact} = 
\delta_{Wdu}(\MW^2)|_{\GW=0} + \delta_{W\nu_l l}(\MW^2)|_{\GW=0}.
\label{eq:fact}
\eeq
Since the vertex corrections for on-shell W~bosons correspond to
physical S-matrix elements, both contributions to the factorizable 
corrections are gauge-invariant. 
Note that $\delta^{\virt}_{\fact}$ is a constant factor, neither
depending on the scattering energy nor on the scattering angle.
Moreover, $\delta^{\virt}_{\fact}$ contains IR singularities originating
from the logarithms of \refeq{eq:onshelllogs}; these terms are connected
to photon emission from on-shell W~bosons and are regularized by the
infinitesimal photon mass $m_\gamma$.

The second category of corrections in the PA are called {\it non-factorizable} 
\cite{Melnikov:1996fx,Denner:1998ia} 
and comprise all remaining resonant contributions,
i.e.\ all terms in $\delta^{\virt}$ that are non-vanishing for $\hat
s\to\MW^2$ and $\GW\to 0$,
\beq
\delta^{\virt}_{\nonfact}(\hat s, \hat t) = 
\delta^{\virt}|_{\hat s\to\MW^2,\GW\to 0} 
- \delta^{\virt}_{\fact}.
\eeq
Here $\delta^{\virt}|_{\hat s\to\MW^2,\GW\to 0}$ results from the full
off-shell correction $\delta^{\virt}$ upon taking the asymptotic
limits $(\hat s-\MW^2)\to 0$ and $\GW\to 0$ while keeping the ratio 
$(\hat s-\MW^2)/(\MW\GW)$ fixed.
As can be shown by simple power counting \cite{Denner:1998ia}, 
only loop diagrams
with an internal photon contribute. The relevant diagrams for
$u\bar d\to\PW\to\nu_l l^+$ are shown in \reffi{fig:nonfact}.
\bfi
\centerline{
\setlength{\unitlength}{1cm}
\begin{picture}(16,7.2)
\put(0,5.0){\includegraphics{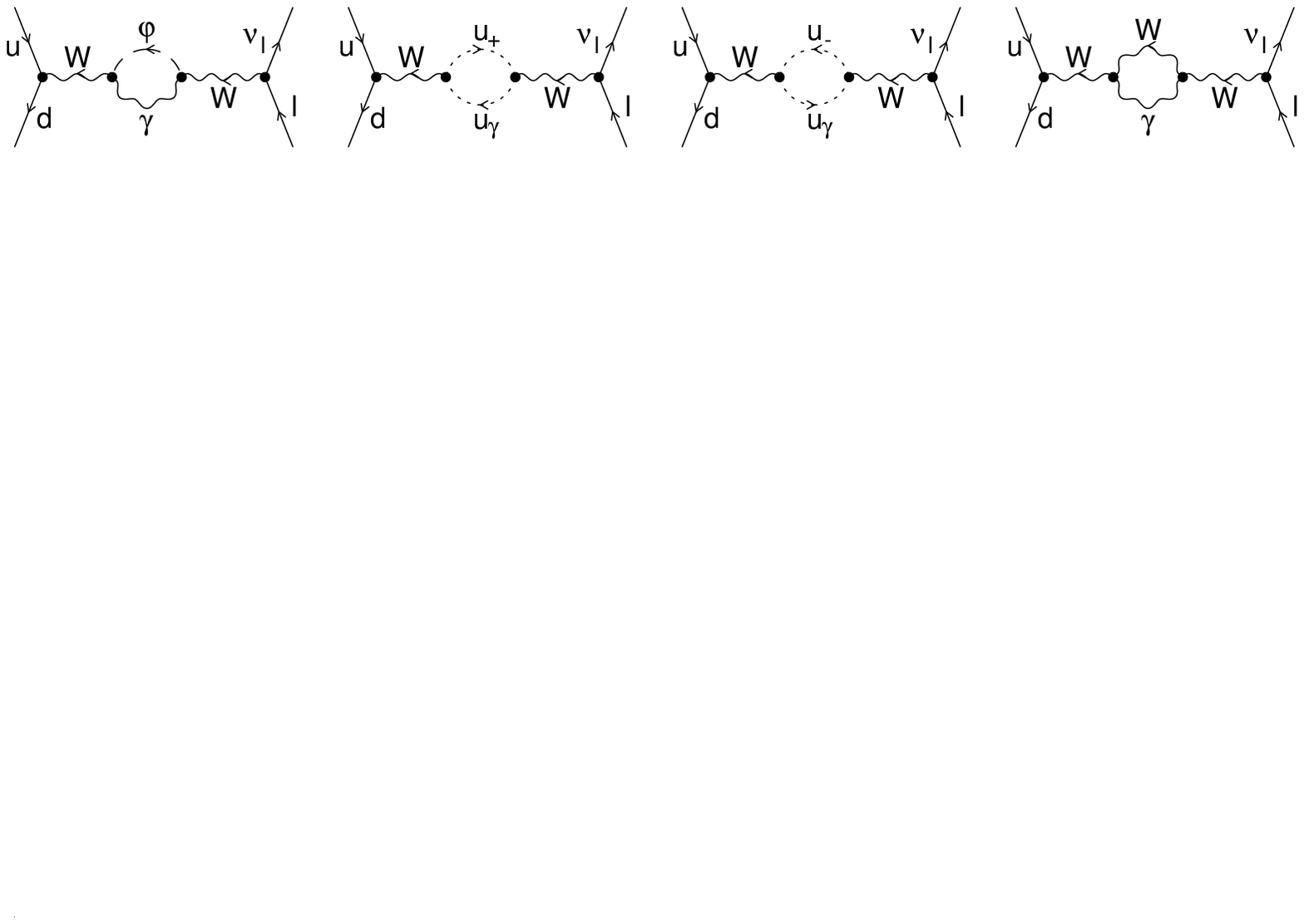}}
\put(0,2.5){\includegraphics{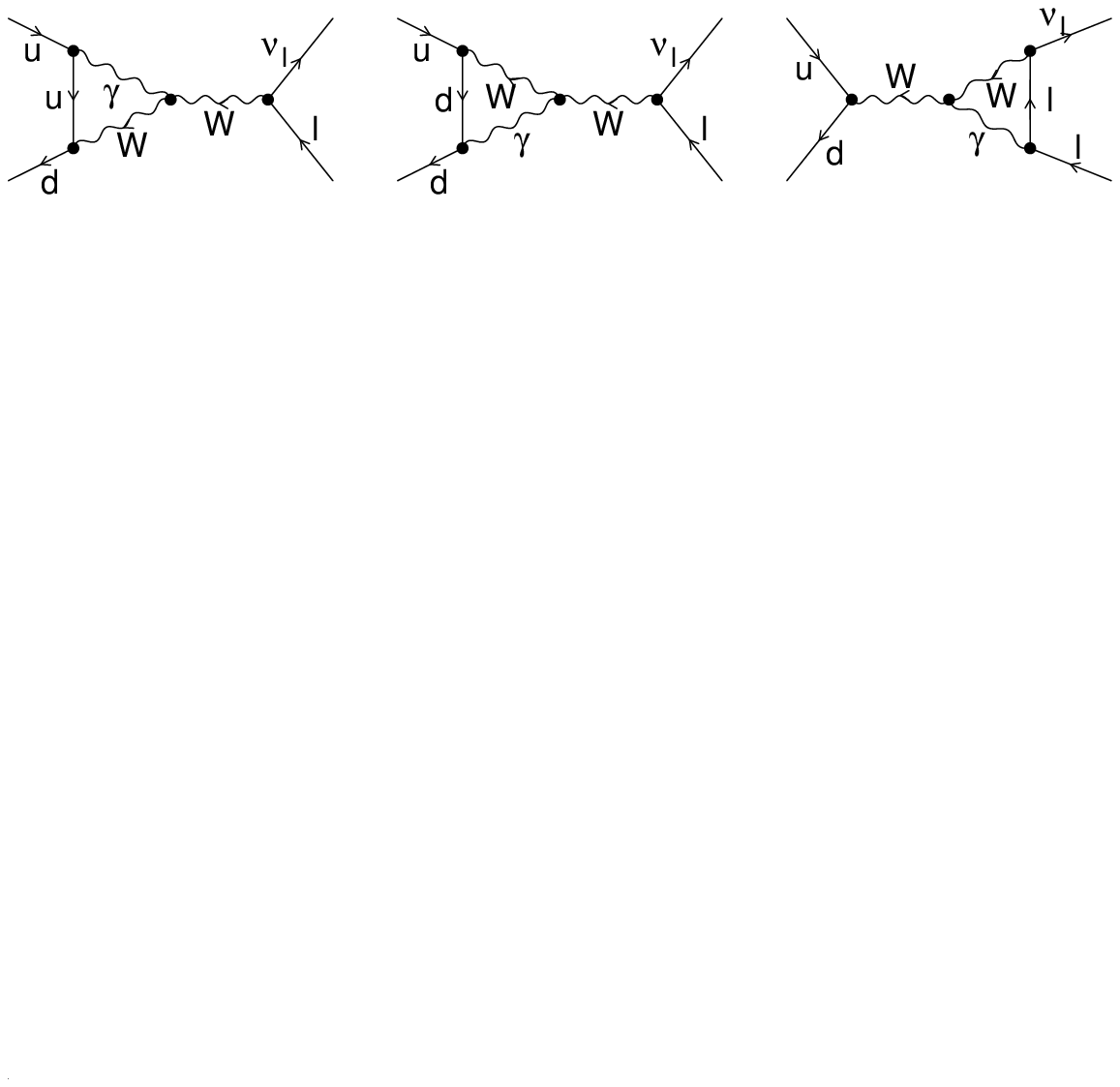}}
\put(0,0){\includegraphics{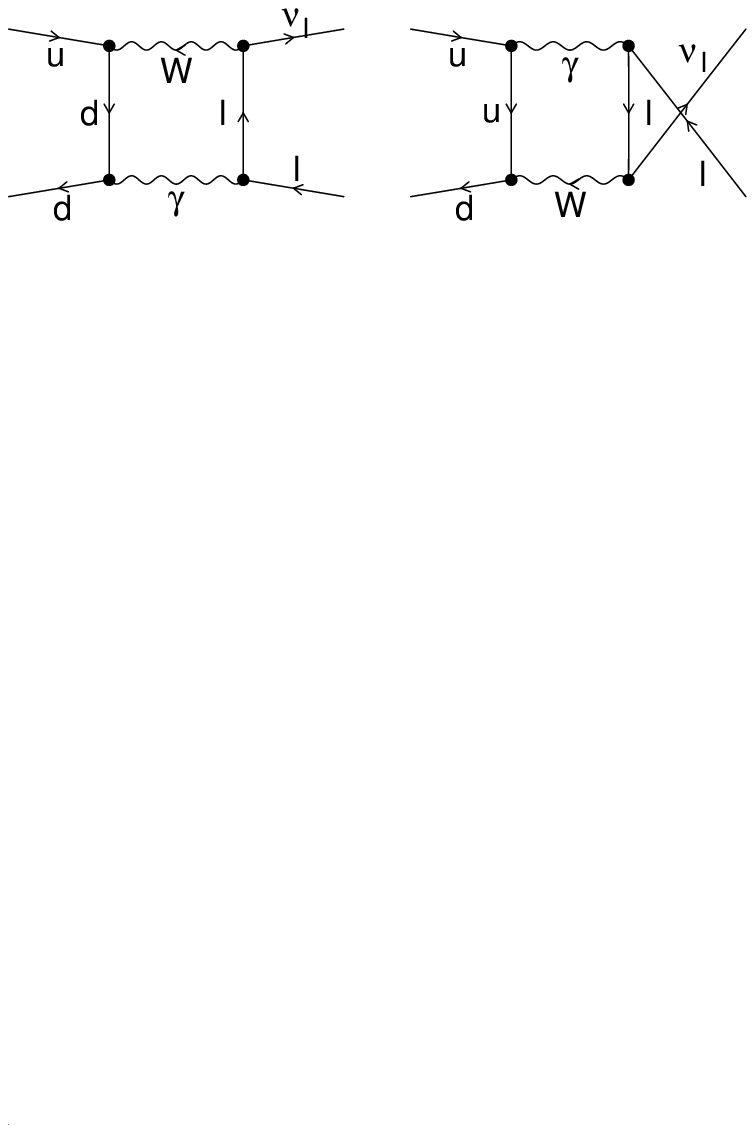}}
\end{picture} 
}
\caption{Diagrams for non-factorizable corrections 
($\varphi$ is the would-be Goldstone partner to the W~boson, and 
$\Pu_\pm$, $\Pu_\gamma$ denote Faddeev--Popov ghosts).}
\label{fig:nonfact}
\efi
Since also box diagrams are involved, production and decay do not
proceed independently, and the terminology {\it non-factorizable} is
justified. The limit $\hat s\to\MW^2$ for the pole expansion has to 
be defined carefully, because $\hat s$ is not the only kinematical
variable. The variables $\hat t$ and $\hat u$, which are related to 
the scattering angle $\hat\theta$ of the parton CM frame, range within 
$-\hat s<\hat t,\hat u<0$ and are related to $\hat s$ by 
$\hat s+\hat t+\hat u=0$. Therefore, changing $\hat s$ while keeping 
$\hat t$ and $\hat u$ fixed is inconsistent in general. 
We circumvent this problem by taking $\hat s\to\MW^2$ for fixed
scattering angle $\hat\theta$, resulting in the replacements
\beq
\hat t \to \hat t_{\res} = \hat t \, \frac{\MW^2}{\hat s} = 
-\MW^2\sin^2\text\frac{\hat\theta}{2}, \qquad
\hat u \to \hat u_{\res} = \hat u \, \disp\frac{\MW^2}{\hat s} =
-\MW^2\cos^2\text\frac{\hat\theta}{2}. 
\eeq
The actual calculation of $\delta^{\virt}_{\nonfact}$ is performed as
described in \citere{Denner:1998ia} in detail. The final result is
\beqar
\lefteqn{
\delta^{\virt}_{\nonfact}(\hat s, \hat t) = 
-\frac{\alpha}{2\pi} \biggl\{
  -2 + Q_d \Li\biggl(1+\frac{\MW^2}{\hat t_{\res}}\biggr)
  - Q_u \Li\biggl(1+\frac{\MW^2}{\hat u_{\res}}\biggr)	}
\nn\\ && \quad {}
  + 2 \ln\biggl(\frac{\MW^2-\ri\MW\GW-\hat s}{m_\gamma\MW}\biggr)
    \biggl[1 + Q_d \ln\biggl(-\frac{\MW^2}{\hat t_{\res}}\biggr) 
             - Q_u \ln\biggl(-\frac{\MW^2}{\hat u_{\res}}\biggr) \biggr] 
\biggr\},
\hspace{2em}
\label{eq:nonfact}
\eeqar
where 
\beq
\Li(x)=-\int_0^x \frac{\rd t}{t} \ln(1-t), \qquad |\arc(1-x)|<\pi,
\eeq
is the usual dilogarithm, and $Q_f$ denotes the electric charge of the
fermion $f$. In \refeq{eq:nonfact} we made use of $Q_l=Q_d-Q_u$.
We note that $\delta^{\virt}_{\nonfact}$ is, by definition, a
gauge-invariant quantity. It does not involve mass singularities of the
fermions, but it is IR-singular. More precisely, the IR-singular term
proportional to $\alpha\ln(m_\gamma)$ exactly compensates the
artificially created IR singularity in $\delta^{\virt}_{\fact}$ by
setting $\hat s=\MW^2$ and $\GW\to 0$ there, and these
$\alpha\ln(m_\gamma)$ terms of $\delta^{\virt}_{\fact}$ are replaced by
the correct logarithms $\alpha\ln(\hat s-\MW^2+\ri\MW\GW)$ after
$\delta^{\virt}_{\nonfact}$ is added.

In summary, the PA for the virtual correction factor reads
\beq
\delta^{\virt}_{\PA} = \delta^{\virt}_{\fact} + 
\delta^{\virt}_{\nonfact}(\hat s, \hat t)
\eeq
with $\delta^{\virt}_{\fact}$ and $\delta^{\virt}_{\nonfact}$ given in
\refeq{eq:fact} and \refeq{eq:nonfact}, respectively. The uncertainty
induced by omitting non-resonant corrections can be estimated naively by
\beq
\delta^{\virt}_{\PA}-\delta^{\virt} \sim 
\frac{\alpha}{\pi}\max\left\{
\frac{\GW}{\MW},
\ln\biggl(\frac{\hat s}{\MW^2}\biggr),
\ln^2\biggl(\frac{\hat s}{\MW^2}\biggr)
\right\},
\label{eq:PAerr}
\eeq
where the factor $\GW/\MW$ results from the neglect of non-resonant
contributions near resonance and the logarithms account for typical
enhancements away from resonance.

Finally, we note that we have analytically compared the PA for the
virtual corrections worked out in this section with the results 
presented in \citere{Wackeroth:1997hz}. Apart from non-resonant
contributions, which go beyond the validity of the PA, both
results agree.%
\footnote{There is a 
 misprint in Eq.~(D.45) in \citere{Wackeroth:1997hz}: the factor 2 in
 front of the $Q_{i'}Q_{f'}[(f,i) \to (f',i')]$, $Q_{i'}Q_{f}[(i,t)
 \to (i',u)]$ and $Q_{i}Q_{f'}[(f,t) \to (f',u)]$ terms needs to be
 removed.}

\subsection{Real-photon emission}

Real-photonic corrections are induced by the diagrams shown in
\reffi{fig:bremdiags}.
\bfi
\centerline{
\setlength{\unitlength}{1cm}
\begin{picture}(16,2.5)
\put(0,0){\includegraphics{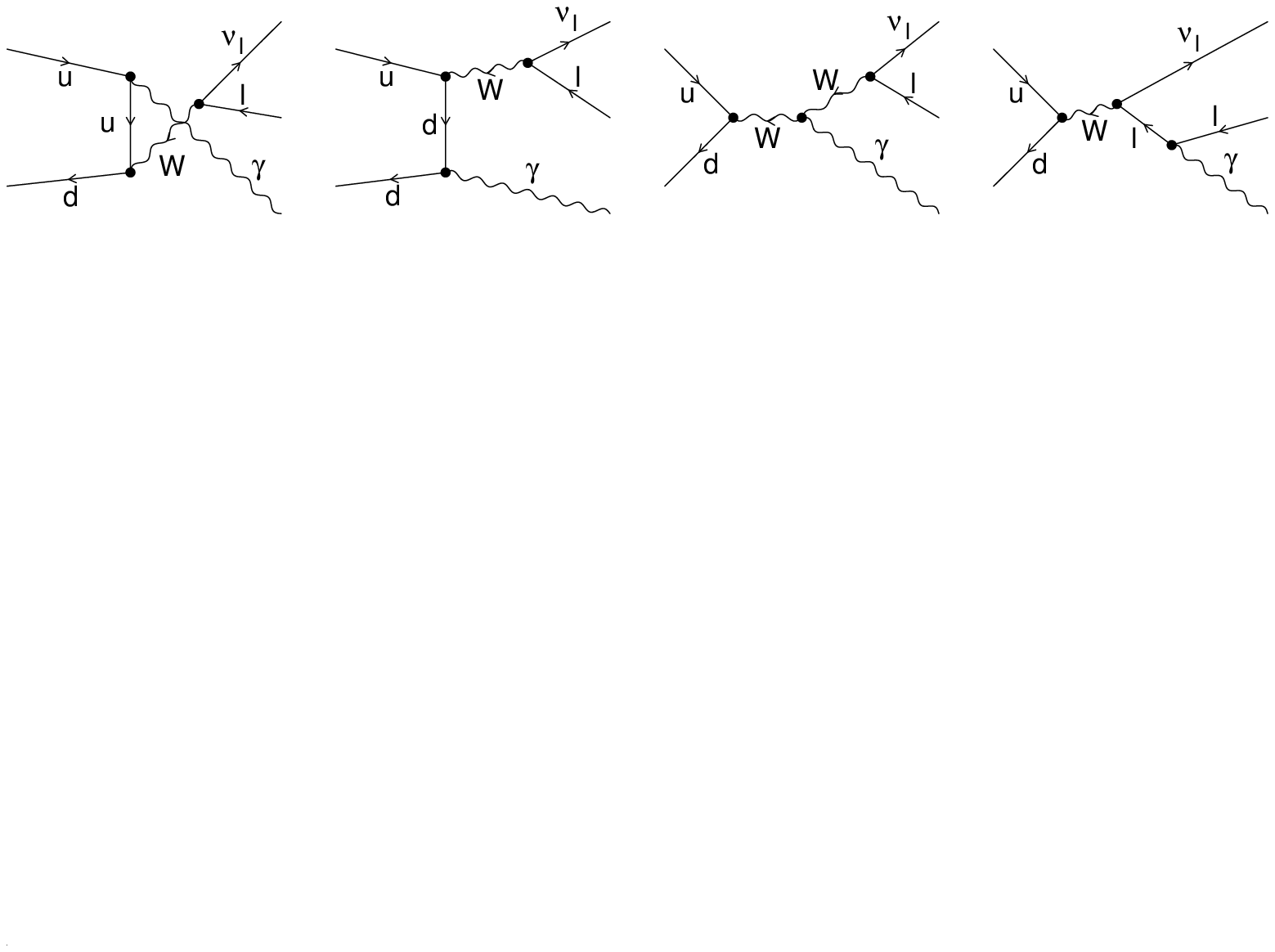}}
\end{picture} 
}
\caption{Diagrams for real-photon emission.}
\label{fig:bremdiags}
\efi
In the calculation of the corresponding amplitudes it is mandatory to
respect the Ward identity for the external photon, i.e.\ electromagnetic
current conservation. Writing the amplitudes as
$\M_\gamma(\lambda)=\veps^*_\mu(k,\lambda) T^\mu$ with $\veps^*_\mu$
denoting the polarization vector of the outgoing photon, 
this Ward identity reads 
$k_\mu T^\mu=0$. If the W~width is zero, this identity is trivially 
fulfilled. This remains even true for a constant width, since the
W-boson mass appears only in the W~propagator denominators, i.e.\ the
substitution $\MW^2\to\MW^2-\ri\MW\GW$ is a consistent
reparametrization of the amplitude in this case.
However, if a running W~width is introduced naively, i.e.\ in the
W~propagators only, the Ward identity is violated.
The Ward identity can be restored by taking into account 
the part of the
fermion-loop correction to the $\gamma WW$ vertex that corresponds to 
the fermion loops in the $W$~self-energy leading to the width in the
propagator \cite{Beenakker:1997kn,Argyres:1995ym}. In \citere{Baur:1995aa} it was shown that this
modification simply amounts to the multiplication of the $\gamma WW$
vertex by the factor
\beq
f_{\gamma WW}\big|_{\mathrm{run}} = 1+\ri\frac{\GW}{\MW}
\eeq
if the photon is on shell ($k^2=0$). 
By construction \cite{Beenakker:1997kn,Argyres:1995ym,Baur:1995aa}, 
the width $\GW$ appearing in $f_{\gamma WW}\big|_{\mathrm{run}}$ as well
as in the W~propagator is the lowest-order width, since it results from 
the imaginary part of the W~self-energy at one loop. Since, however, 
for the special case of an on-shell photon the relevant imaginary parts
are completely parametrized by the ratio $\GW/\MW$, the Ward identity is
fulfilled for any numerical value of $\GW$. Therefore, we are allowed
to use a value for $\GW$ that includes also QCD and electroweak
radiative corrections.
For later convenience, we define
\beq
f_{\gamma WW}\big|_{\mathrm{const}} = 1.
\eeq

In addition to the U(1) Ward identity for the external photon,
an SU(2) Ward identity \cite{Beenakker:1997kn,Argyres:1995ym}
becomes relevant for effectively longitudinally polarized W~bosons 
in $\PW\gamma$ production at high energies, where the lepton current 
$[\bar u_{\nu_l}\gamma_\mu\omega_-v_l]$ becomes proportional to the
W~momentum $k_\PW=k_n+k_l$. As can be checked easily, this Ward
identity is maintained in the above treatment of the W~width as well.

The helicity amplitudes $\M^{\tau_u\tau_d\tau_l}_\gamma(\lambda)$ 
for the radiative process 
$u\bar d\to\nu_l l^+\gamma$ can be written in a very compact way using
the Weyl-van der Waerden spinor formalism. Adopting the conventions of
\citere{Dittmaier:1999nn}, we obtain
\newcommand{\pipii}{\langle p_u p_d\rangle}
\newcommand{\piiki}{\langle p_d k_n\rangle}
\newcommand{\pikii}{\langle p_u k_l\rangle}
\newcommand{\pik}{\langle p_u k\rangle}
\newcommand{\piik}{\langle p_d k\rangle}
\newcommand{\kki}{\langle k k_n\rangle}
\newcommand{\kkii}{\langle k k_l\rangle}
\newcommand{\kikii}{\langle k_n k_l\rangle}
\newcommand{\Cpipii}{\pipii^*}
\newcommand{\Cpikii}{\pikii^*}
\newcommand{\Cpik}{\pik^*}
\newcommand{\Cpiik}{\piik^*}
\newcommand{\Ckki}{\kki^*}
\newcommand{\Ckkii}{\kkii^*}
\newcommand{\Ckikii}{\kikii^*}
\beqar
\M^{-+-}_\gamma(+1) &=& \frac{\sqrt{2}e^3 V^*_{ud}}{\sw^2} \piiki^2 \biggl\{ 
-\frac{Q_u\Ckikii}{[s_{\nu_l l}-\MW^2+\ri\MW\GW(s_{\nu_l l})]\pik\piik} 
\nn\\
&& \quad
{}+\frac{1}{\hat s-\MW^2+\ri\MW\GW(\hat s)}
  \biggl[ \frac{Q_l\Cpipii}{\kki\kkii}
         +\frac{(Q_d-Q_u)\Cpikii}{\piik\kki} \biggr]
\nn\\
&& \quad
{}+\frac{f_{\gamma\PW\PW}(Q_u-Q_d)\Ckikii\Cpik}
       {[\hat s-\MW^2+\ri\MW\GW(\hat s)]
	[s_{\nu_l l}-\MW^2+\ri\MW\GW(s_{\nu_l l})]\piik}  \biggr\},
\nn\\[.5em]
\M^{-+-}_\gamma(-1) &=& \frac{\sqrt{2}e^3 V^*_{ud}}{\sw^2} (\Cpikii)^2 \biggl\{ 
-\frac{Q_d\kikii}{[s_{\nu_l l}-\MW^2+\ri\MW\GW(s_{\nu_l l})]\Cpik\Cpiik} 
\nn\\
&& \quad
{}+\frac{1}{\hat s-\MW^2+\ri\MW\GW(\hat s)}
  \biggl[ \frac{Q_{\nu_l}\pipii}{\Ckki\Ckkii}
         +\frac{(Q_d-Q_u)\piiki}{\Cpik\Ckkii} \biggr]
\nn\\
&& \quad
{}+\frac{f_{\gamma\PW\PW}(Q_d-Q_u)\kikii\piik}
       {[\hat s-\MW^2+\ri\MW\GW(\hat s)]
	[s_{\nu_l l}-\MW^2+\ri\MW\GW(s_{\nu_l l})]\Cpik}  \biggr\}.
\hspace{2em}
\label{eq:mreal}
\eeqar
The amplitudes for the other helicity channels vanish for massless fermions.
The spinor products are defined by
\beq
\langle pq\rangle=\epsilon^{AB}p_A q_B
=2\sqrt{p_0 q_0} \,\Biggl[
{\mathrm{e}}^{-\ri\phi_p}\cos\frac{\theta_p}{2}\sin\frac{\theta_q}{2}
-{\mathrm{e}}^{-\ri\phi_q}\cos\frac{\theta_q}{2}\sin\frac{\theta_p}{2}
\Biggr],
\eeq
where $p_A$, $q_A$ are the associated momentum spinors for the light-like
momenta
\beqar
p^\mu&=&p_0(1,\sin\theta_p\cos\phi_p,\sin\theta_p\sin\phi_p,\cos\theta_p),\nl
q^\mu&=&p_0(1,\sin\theta_q\cos\phi_q,\sin\theta_q\sin\phi_q,\cos\theta_q).
\eeqar

The contribution $\hat\sigma_\gamma$ of the radiative process to the
parton cross section is given by
\beq
\hat\sigma_\gamma = \frac{1}{12}\frac{1}{2s} \int \rd\Gamma_\gamma \,
\sum_{\lambda} |\M^{-+-}_\gamma(\lambda)|^2,
\label{eq:hbcs}
\eeq
where the phase-space integral is defined by
\beq
\int \rd\Gamma_\gamma =
\int\frac{\rd^3 {\bf k}_n}{(2\pi)^3 2k_{n,0}}
\int\frac{\rd^3 {\bf k}_l}{(2\pi)^3 2k_{l,0}}
\int\frac{\rd^3 {\bf k}}{(2\pi)^3 2k_0} \,
(2\pi)^4 \delta(p_u+p_d-k_n-k_l-k).
\label{eq:dGg}
\eeq

\subsection{Treatment of soft and collinear singularities}
\label{se:virt+real}

The phase-space integral \refeq{eq:hbcs} diverges in the soft ($k_0\to
0$) and collinear ($p_u k, p_d k, k_l k\to 0$) regions logarithmically
if the photon and fermion masses are set to zero, as done in
\refeq{eq:mreal}. For the treatment of the soft and collinear
singularities we applied three different methods, the results of which
are in good numerical agreement. In the following we briefly sketch
these approaches.

\subsubsection{IR phase-space slicing and effective collinear factors}

Firstly, we
made use of the variant of phase-space slicing that is described in 
\citere{Berends:1982uq}, where the soft-photon region is excluded in the
integral \refeq{eq:hbcs} but the regions of photon emission
collinear to the fermions are included.

In the soft-photon region $m_\gamma<k_0<\Delta E\ll\sqrt{\hat s}$
the bremsstrahlung cross section factorizes into the lowest-order 
cross section and a universal eikonal factor that depends on the
photon momentum $k$ (see, e.g., \citere{Denner:1993kt}). Integration
over $k$ in the partonic CM frame yields a simple correction factor 
$\delta_{\soft}$ to the partonic Born cross section $\rd\hat\sigma_0$,
\beqar
\delta_{\soft} &=&
-\frac{\alpha}{2\pi}\Biggl\{
 Q_l^2 \left[ 
   2\ln\biggl(\frac{2\Delta E}{m_\gamma}\biggr)
        \ln\biggl(\frac{m_l^2}{\hat s}\biggr)
   +2\ln\biggl(\frac{2\Delta E}{m_\gamma}\biggr) 
   +\frac{1}{2}\ln^2\biggl(\frac{m_l^2}{\hat s}\biggr)
   +\ln\biggl(\frac{m_l^2}{\hat s}\biggr) 
   +\frac{\pi^2}{3}
	\right]
\nn\\ && \hspace*{2.5em} {}
+Q_d^2 \left[ 
   2\ln\biggl(\frac{2\Delta E}{m_\gamma}\biggr)
        \ln\biggl(\frac{m_d^2}{\hat s}\biggr)
   +2\ln\biggl(\frac{2\Delta E}{m_\gamma}\biggr) 
   +\frac{1}{2}\ln^2\biggl(\frac{m_d^2}{\hat s}\biggr)
   +\ln\biggl(\frac{m_d^2}{\hat s}\biggr) 
   +\frac{\pi^2}{3}
	\right]
\nn\\ && \hspace*{2.5em} {}
+Q_u^2 \left[ 
   2\ln\biggl(\frac{2\Delta E}{m_\gamma}\biggr)
        \ln\biggl(\frac{m_u^2}{\hat s}\biggr)
   +2\ln\biggl(\frac{2\Delta E}{m_\gamma}\biggr) 
   +\frac{1}{2}\ln^2\biggl(\frac{m_u^2}{\hat s}\biggr)
   +\ln\biggl(\frac{m_u^2}{\hat s}\biggr) 
   +\frac{\pi^2}{3}
	\right]
\nn\\ && \hspace*{2.5em} {}
-2Q_l Q_d \left[ 2\ln\biggl(\frac{2\Delta E}{m_\gamma}\biggr)
	\ln\biggl(\frac{-\hat t}{\hat s}\biggr)
                -\Li\biggl(-\frac{\hat u}{\hat t}\biggr) \right]
\nn\\ && \hspace*{2.5em} {}
+2Q_l Q_u \left[ 2\ln\biggl(\frac{2\Delta E}{m_\gamma}\biggr)
	\ln\biggl(\frac{-\hat u}{\hat s}\biggr)
                -\Li\biggl(-\frac{\hat t}{\hat u}\biggr) \right]
\Biggr\}. 
\label{eq:dsoft}
\eeqar
The factor $\delta_{\soft}$ can be added directly to the virtual
correction factor $2\Re\{\delta^{\virt}\}$ defined in
\refeq{eq:dm2virt}. It can be checked easily that the photon mass 
$m_\gamma$ cancels in the sum $2\Re\{\delta^{\virt}\}+\delta_{\soft}$.

The remaining phase-space integration in \refeq{eq:hbcs} with
$k_0>\Delta E$ still contains the collinear singularities in the
regions in which $(p_u k)$, $(p_d k)$, or $(k_l k)$ is small. 
In these regions,
however, the asymptotic behaviour of the differential cross section
(including its dependence on the fermion masses) has a well-known
form. The singular terms are universal and factorize from
$\rd\sigma_0$. A simple approach to include the collinear regions
consists in a suitable modification of $|\M_\gamma|^2$, which was
calculated for vanishing fermion masses. More precisely,
$|\M_\gamma|^2$ is multiplied by an {\it effective collinear factor}
that is equal to 1 up to terms of ${\cal O}(m_f^2/\hat s)$ ($f=l,u,d$)
outside the collinear regions, but replaces the poles in $(p_u k)$,
$(p_d k)$, and
$(k_l k)$ by the correctly mass-regularized behaviour.  Explicitly,
the substitution reads
\beqar
\sum_{\lambda=\pm 1} |\M_\gamma^{\tau_u,\tau_d,\tau_l}(\lambda)|^2
& \;\to\; &
\sum_{\kappa_u,\kappa_d,\kappa_l=\pm 1}
f_{\kappa_u}^{(\ini)}(m_u,x_u,E_u,\theta_{u\gamma}) 
f_{\kappa_d}^{(\ini)}(m_d,x_d,E_d,\theta_{d\gamma})
\nn\\
&& {} \times 
f_{\kappa_l}^{(\fin)}(m_l,x_l,E_l,\theta_{l\gamma})
\sum_{\lambda=\pm 1}
|\M_\gamma^{\kappa_u\tau_u,\kappa_d\tau_d,\kappa_l\tau_l}(\lambda)|^2.
\eeqar
The functions $f^{(\ini/\fin)}_+$ and $f^{(\ini/\fin)}_-$ 
describe collinear photon emission with and
without spin flip of the radiating fermion, respectively,
\beqar
f^{(\ini/\fin)}_+(m_f,x_f,E_f,\theta_{f\gamma}) & = &
\left(\frac{4E_f^2\sin^2({\theta_{f\gamma}}/{2})}
{4E_f^2\sin^2({\theta_{f\gamma}}/{2})+\Mf^2}\right)^2,
\nn\\
f^{(\ini/\fin)}_-(m_f,x_f,E_f,\theta_{f\gamma}) & = &
\frac{x_f^2}{x_f^2\mp 2x_f+2} \frac{4\Mf^2 E_f^2\sin^2({\theta_{f\gamma}}/{2})}
{[4E_f^2\sin^2({\theta_{f\gamma}}/{2})+\Mf^2]^2},
\qquad x_f=\frac{k_0}{E_f},
\hspace{2em}
\eeqar
where $E_f$ is the fermion energy and
$\theta_{f\gamma}=\angle({\bf k}_f,{\bf k})$ is the angle of the photon
emission from $f=u,d,l$.

\subsubsection{IR and collinear phase-space slicing}

Instead of using effective collinear factors, alternatively
we have also applied
phase-space slicing to the collinear singularities, i.e.\ the collinear
regions are now excluded by the angular cuts
$\theta_{f\gamma}<\Delta\theta\ll 1$ in the integral \refeq{eq:hbcs}.
The IR region is treated as previously, leading to the same correction
factor $\delta_{\soft}$ as given in \refeq{eq:dsoft}.

In the collinear cones the photon emission angles $\theta_{f\gamma}$
can be integrated out by making use of the factorization property
of the squared photon-emission matrix elements 
with the radiator functions $f^{(\ini/\fin)}_\pm$, as described in
the previous section. The resulting contribution to the 
bremsstrahlung cross section has the form of a convolution of the
lowest-order cross section,
\beqar
\sigma_{\coll} &=& \sigma_{\coll,u} + \sigma_{\coll,d} + \sigma_{\coll,l}, 
\\[.5em]
\sigma_{\coll,q}(p_q) &=&
\frac{Q_q^2 \alpha}{2\pi} \int_0^{1-2\Delta E/\sqrt{\hat s}} \rd z\,
\Biggl\{ \left[ \ln\biggl(\frac{\Delta\theta^2 \hat s}{4m_q^2}\biggr)-1\right]
P_{ff}(z) \sigma_0(zp_q)
\nn\\
&& \hspace*{9em} {}
+ (1-z)\sigma_0(zp_q)\Big|_{\tau_q\to-\tau_q} \Biggr\},
\qquad q=u,d,
\label{eq:qcoll}
\\[.5em]
\sigma_{\coll,l}(k_l) &=&
\frac{Q_l^2 \alpha}{2\pi} \int_0^{1-2\Delta E/\sqrt{\hat s}} \rd z\,
\Biggl\{ 
\left[ 
\ln\biggl(\frac{\Delta\theta^2 \hat s}{4m_l^2}\biggr)+2\ln(z)-1\right]
P_{ff}(z) \sigma_0(k_l)
\nn\\
&& \hspace*{9em} {}
+ (1-z)\sigma_0(k_l)\Big|_{\tau_l\to-\tau_l} \Biggr\},
\label{eq:lcoll}
\eeqar
with the splitting function
\beq
P_{ff}(z) = \frac{1+z^2}{1-z}.
\eeq
For initial-state radiation the respective quark momentum $p_q$ is
reduced by the factor $z$ so that the partonic CM frame for the hard scattering
receives a boost, while this is not the case for final-state radiation.
Note that for final-state radiation, i.e.\ in $\sigma_{\coll,l}$,
the lepton momentum in the final state is $zk_l$, although $k_l$ is
relevant for the lowest-order cross section in the hard scattering.
Of course, if photons collinear to the lepton $l$ are not separated the
$z$ integration can be carried out explicitly. It can be checked
easily that in this case all logarithms of the lepton mass $m_l$
cancel in the sum of virtual and real corrections.

\subsubsection{Subtraction method}
\label{se:sub}

Finally, we applied the subtraction method presented in 
\citere{Dittmaier:2000mb}, where the so-called ``dipole formalism'', 
originally introduced by Catani and Seymour \cite{Catani:1996jh}
within massless QCD, was applied to photon radiation and generalized
to massive fermions.
The general idea of a subtraction method is to subtract and
to add a simple auxiliary function from the singular integrand.
This auxiliary function has to be chosen such that it cancels all
singularities of the original integrand so that the phase-space
integration of the difference can be performed numerically.
Moreover, the auxiliary function has to be simple enough so that it can
be integrated over the singular regions analytically, when the
subtracted contribution is added again.

The dipole subtraction function consists of contributions
labelled by all ordered pairs of charged external particles, one of
which is called {\it emitter}, the other one {\it spectator}.
For $u\bar d\to\nu_l l^+$ we, thus, have six different 
emitter/spectator cases $ff'$: $ud$, $du$, $ul$, $lu$, $dl$, $ld$.
The subtraction function that is subtracted from 
$\sum_{\lambda}|\M_\gamma(\lambda)|^2$ is given by
\beqar
|\M_{\sub}|^2 &=& \phantom{{}+{}}
Q_u Q_d e^2 \Bigl[ g_{ud}(p_u,p_d,k) |\M_\born(x_{ud}p_u,p_d,k_{l,ud})|^2
\nn\\
&& \hspace*{4.5em} {}
+g_{du}(p_d,p_u,k) |\M_\born(p_u,x_{ud}p_d,k_{l,du})|^2 \Bigr]
\nn\\
&& {}
- Q_u Q_l e^2 \Bigl[ g_{ul}(p_u,k_l,k) + g_{lu}(k_l,p_u,k) \Bigr]
|\M_\born(x_{ul}p_u,p_d,k_{l,ul})|^2 
\nn\\
&& {}
+ Q_d Q_l e^2 \Bigl[ g_{dl}(p_d,k_l,k) + g_{ld}(k_l,p_d,k) \Bigr]
|\M_\born(p_u,x_{dl}p_d,k_{l,dl})|^2, 
\label{eq:msub}
\eeqar
with the functions
\beqar
g_{ud}(p_u,p_d,k) &=& \frac{1}{(p_u k)x_{ud}}
\left[ \frac{2}{1-x_{ud}}-1-x_{ud} \right],
\nn\\
g_{du}(p_d,p_u,k) &=& \frac{1}{(p_d k)x_{ud}}
\left[ \frac{2}{1-x_{ud}}-1-x_{ud} \right],
\nn\\
g_{ql}(p_q,k_l,k) &=& \frac{1}{(p_q k)x_{ql}}
\left[ \frac{2}{2-x_{ql}-z_{ql}}-1-x_{ql} \right],
\nn\\
g_{lu}(k_l,p_q,k) &=& \frac{1}{(k_l k)x_{ql}}
\left[ \frac{2}{2-x_{ql}-z_{ql}}-1-z_{ql} \right], \qquad q=u,d,
\eeqar
and the auxiliary variables
\beq
x_{ud} = \frac{p_u p_d-p_u k-p_d k}{p_u p_d}, \qquad
x_{ql} = \frac{p_q k_l +p_q k-k_l k}{p_q k_l +p_q k}, \qquad
z_{ql} = \frac{p_q k_l }{p_q k_l +p_q k}.
\eeq
For the evaluation of $|\M_{\sub}|^2$ in \refeq{eq:msub} the lepton
momenta $k_{l,ff'}$ still have to be specified.
They are given by
\beq
k_{l,ud}^\mu = {\Lambda(p_u,p_d)^\mu}_\nu k_l^\nu, \qquad
k_{l,du}^\mu = {\Lambda(p_d,p_u)^\mu}_\nu k_l^\nu, \qquad
k_{l,ql}^\mu = k_l^\mu+k^\mu-(1-x_{ql})p_q^\mu,
\eeq
with the Lorentz transformation matrix
\beq
{\Lambda(p_1,p_2)^\mu}_\nu = 
{g^\mu}_\nu - \frac{(P+\tilde P)^\mu(P+\tilde P)_\nu}{P^2+P\tilde P}
+\frac{2\tilde P^\mu P_\nu}{P^2}, \qquad
P^\mu=k_l^\mu+k_n^\mu, \quad \tilde P^\mu=x_{ud}p_1^\mu+p_2^\mu.
\eeq
The modified lepton momenta $k_{l,ff'}$ still obey the on-shell
condition $k_{l,ff'}^2=0$, and the same is true for the corresponding
neutrino momenta that result from momentum conservation.  It is
straightforward to check that all collinear and soft singularities
cancel in $\sum_{\lambda}|\M_\gamma(\lambda)|^2-|\M_{\sub}|^2$ so that
this difference can be integrated numerically over the entire phase
space \refeq{eq:dGg}. If phase-space cuts are applied to the lepton
momentum, these cuts directly affect $k_l$ in
$\sum_{\lambda}|\M_\gamma(\lambda)|^2$, but in $|\M_{\sub}|^2$ they
have to be applied to $k_{l,ff'}$. The singularities nevertheless
properly cancel in this case, since the momenta $k_{l,ff'}$ are
defined in such a way that they asymptotically approach $k_l$ in the
singular regions.

The contribution of $|\M_{\sub}|^2$, which has been subtracted by
hand, has to be added again. This is done after the singular degrees
of freedom in the phase space \refeq{eq:dGg} are integrated out
analytically, keeping an infinitesimal photon mass $m_\gamma$ and
small fermion masses $m_f$ as regulators \cite{Dittmaier:2000mb}.  The
resulting contribution is split into two parts: one that factorizes
from the lowest-order cross section $\sigma_0$ and another part that
has the form of a convolution integral over $\sigma_0$ with reduced CM
energy.  The first part is given by
\beqar
\sigma_{\sub,1} &=& 
\frac{\alpha}{2\pi} \Biggl\{ 
Q_u Q_d\left[ {\cal L}( \hat s,m_u^2,m_d^2) 
	+3-\frac{2\pi^2}{3} \right]
- Q_l Q_u \left[ {\cal L}(-\hat u,m_l^2,m_u^2) 
	-\frac{1}{2}-\frac{\pi^2}{3}\right]
\nn\\ && \qquad {}
+ Q_l Q_d \left[ {\cal L}(-\hat t,m_l^2,m_d^2) 
	-\frac{1}{2}-\frac{\pi^2}{3}\right]
\Biggr\} \sigma_0
+\frac{Q_l^2\alpha}{4\pi} \sigma_0\Big|_{\tau_l\to -\tau_l}
\eeqar
with the auxiliary function
\beq
{\cal L}(r,m_1^2,m_2^2) =
2\ln\biggl(\frac{m_1 m_2}{r}\biggr)
\ln\biggl(\frac{m_\gamma^2}{r}\biggr)
+ 2\ln\biggl(\frac{m_\gamma^2}{r}\biggr)
- \frac{1}{2}\ln^2\biggl(\frac{m_1^2}{r}\biggr)
- \frac{1}{2}\ln^2\biggl(\frac{m_2^2}{r}\biggr)
+ \ln\biggl(\frac{m_1 m_2}{r}\biggr).
\label{eq:L}
\eeq
The IR and fermion-mass singularities contained in 
$\rd\sigma_{\sub,1}$ exactly cancel the ones of the virtual corrections.
The second integrated subtraction contribution is given by%
\footnote{Note that we did not literally follow the formulae of
\citere{Dittmaier:2000mb}, but rearranged some terms in the spin-flip parts.}
\beqar
\sigma_{\sub,2}(p_u,p_d) &=&
Q_u Q_d \frac{\alpha}{2\pi} \int_0^1 \rd x\, \left\{
 \left[ {\cal G}_{ud}(\hat s,x) \right]_+ \sigma_0(xp_u,p_d)
\right.
\nn\\ && \left. \qquad\qquad\qquad\qquad {}
+\left[ {\cal G}_{du}(\hat s,x) \right]_+ \sigma_0(p_u,xp_d) \right\}
\Big|_{\hat s=(p_u+p_d)^2}
\nn\\ && {} 
-Q_l Q_u \frac{\alpha}{2\pi} \int_{-\hat s}^0 \rd\hat u\, \int_0^1 \rd x\, 
\left[ {\cal G}_{lu}(-\hat u,x) \right]_+ 
\frac{\rd\sigma_0}{\rd\hat u}(xp_u,p_d)
\Big|_{\hat u=(p_d-k_n)^2}
\nn\\ && {} 
+Q_l Q_d \frac{\alpha}{2\pi} \int_{-\hat s}^0 \rd\hat t\, \int_0^1 \rd x\, 
\left[ {\cal G}_{ld}(-\hat t,x) \right]_+ 
\frac{\rd\sigma_0}{\rd\hat t}(p_u,xp_d)
\Big|_{\hat t=(p_u-k_n)^2}
\nn\\ && {} 
+\frac{\alpha}{2\pi} \int_0^1 \rd x\, (1-x)
\left\{ Q_u^2\sigma_0(xp_u,p_d)\Big|_{\tau_u\to -\tau_u}
        +Q_d^2\sigma_0(p_u,xp_d)\Big|_{\tau_d\to -\tau_d} \right\},
\nn\\
\label{eq:subconv}
\eeqar
where the usual $[\dots]_+$ prescription,
\beq
\int_0^1\rd x\, \Big[f(x)\Big]_+ g(x) =
\int_0^1\rd x\, f(x) \left[g(x)-g(1)\right],
\eeq
is applied to the integration kernels
\beqar
{\cal G}_{qq'}(r,x) &=& 
P_{ff}(x) \left[ \ln\left(\frac{r}{m_q^2}\right)-1 \right],
\nn\\
{\cal G}_{lq}(r,x) &=& 
P_{ff}(x) \left[ \ln\left(\frac{r}{x(1-x)m_q^2}\right)-1 \right]
-\frac{3}{2(1-x)}.
\eeqar
In \refeq{eq:subconv} we indicated explicitly how the
Mandelstam variable $r$ has to be chosen in terms of the momenta
in the evaluation of the part containing $[{\cal G}_{ff'}(r,x)]_+$.
Note, however, that in \refeq{eq:subconv} the variable $\hat s$
that is implicitly used in the calculation of $\sigma_0(\dots)$
is reduced to $2xp_u p_d$.

In summary, within the subtraction approach the real correction reads
\beq
\hat\sigma_\gamma = \frac{1}{12}\frac{1}{2s} \int \rd\Gamma_\gamma \,
\left[ \sum_{\lambda} |\M_\gamma(\lambda)|^2-|\M_{\sub}|^2 \right]
+ \sigma_{\sub,1} + \sigma_{\sub,2}.
\label{eq:sigmasub}
\eeq
It should be realized that in $\sigma_{\sub,1}$ and $\sigma_{\sub,2}$
the full photonic phase space is integrated over. This does, however,
not restrict the subtraction approach to observables that are fully
inclusive with respect to emitted photons, but rather to observables that 
are inclusive with respect to photons that are soft or collinear to any
charged external fermion (see discussions in Sect.~6.2 of 
\citere{Dittmaier:2000mb} and Sect.~7 of \citere{Catani:1996jh}).
In this context, the implementation of
phase-space cuts, which define the observable, in Eq.~\refeq{eq:sigmasub} 
is crucial. 
The phase-space cuts have to be applied in the subtraction parts 
$\sigma_{\sub,1}$, $\sigma_{\sub,2}$, and $|\M_{\sub}|^2$, which are all
defined on the $2\to 2$ kinematics, exactly in the same way, in order
to guarantee that these parts consistently cancel each other.
On the other hand, the observable-defining cuts directly enter the
$2\to 3$ part $\sum_{\lambda} |\M_\gamma(\lambda)|^2$.
Consistency requires that these cuts treat soft or collinear photons
inclusively, i.e.\ that photons collinear to charged fermions have to be
associated with these fermions; otherwise $|\M_{\sub}|^2$ cannot locally 
cancel the collinear singularities in 
$\sum_{\lambda} |\M_\gamma(\lambda)|^2$.

\section{The hadronic processes 
\boldmath{$\Pp\Pp,\Pp\bar\Pp\to\PWp\to l^+\nu_l(+\gamma)$}}
\label{se:ppcs}

The proton--(anti-)proton cross section $\sigma$ is obtained from the
parton cross sections $\hat\sigma^{(q_1 q_2)}$ by convolution with the
corresponding parton distribution functions $q_{1,2}(x)$,
\beq
\sigma(s) = \sum_{q_1 q_2} \int_0^1 \rd x_1 \, \int_0^1 \rd x_2 \,
q_1(x_1) q_2(x_2) \hat\sigma^{(q_1 q_2)}(x_1 p_{q_1}, x_2 p_{q_2}).
\label{eq:sigpp}
\eeq
In the sum $\sum_{q_1 q_2}$ the quark pairs $q_1 q_2$ run over all
possible combinations $u\bar d$ and $\bar d u$ of up-type quarks
$u=\Pu,\Pc$ and down-type quarks $d=\Pd,\Ps$. The squared CM energy $s$
of the $\Pp\Pp$ ($\Pp\bar\Pp$) system is related to the squared parton
CM energy $\hat s$ by $\hat s=x_1 x_2 s$.

The ${\cal O}(\alpha)$-corrected parton cross section
$\hat\sigma^{(q_1 q_2)}$ contains mass singularities of the form
$\alpha\ln(m_q)$, which are due to collinear photon radiation off the
initial-state quarks.  In complete analogy to the
$\overline{\mbox{MS}}$ factorization scheme for next-to-leading order
QCD corrections, we absorb these collinear singularities into the
quark distributions.  This is achieved by replacing $q(x)$ in
\refeq{eq:sigpp} according to 
\beq
q(x) \to q(x,M^2) -
\int_x^1 \frac{\rd z}{z} \, q\biggl(\frac{x}{z},M^2\biggr) \,
\frac{\alpha}{2\pi} \, Q_q^2 \, \biggl[ P_{ff}(z)
\biggl\{\ln\biggl(\frac{M^2}{m_q^2}\biggr)-2\ln(1-z)-1\biggr\} \biggr]_+ ,
\label{eq:factorization}
\eeq
where $M$ is the factorization scale. 
This replacement defines the
same finite parts in the ${\cal O}(\alpha)$ correction as the usual 
$\overline{\mbox{MS}}$ factorization in $D$-dimensional regularization
for exactly massless partons, where the $\ln(m_q)$ terms appear as
$1/(D-4)$ poles.
In \refeq{eq:factorization}
we have preferred to exclude the
soft-photon pole by using the $[\dots]_+$
prescription. This procedure is fully equivalent to the application of
a soft-photon cutoff, as used in \citere{Baur:1999kt}.

The absorption of the collinear singularities of ${\cal O}(\alpha)$
into quark distributions, as a matter of fact, requires also the
inclusion of the corresponding ${\cal O}(\alpha)$ corrections into the
DGLAP evolution of these distributions and into their fit to
experimental data. At present, this full incorporation of ${\cal
O}(\alpha)$ effects in the determination of the quark distributions
has not yet been performed. However, an approximate inclusion of the
${\cal O}(\alpha)$ corrections to the DGLAP evolution shows
\cite{Kripfganz:1988bd} that the impact of these corrections on the
quark distributions is well below 1\%, at least in the $x$ range that
is relevant for W-boson production at the Tevatron and the LHC.
Therefore, the neglect of these corrections to the parton distributions
is justified for the following numerical study.

\section{Numerical results}
\label{se:numres}

\subsection{Input parameters}

For the numerical evaluation we used the following set of parameters,
\beq
\begin{array}[b]{lcllcllcl}
\alpha &=& 1/137.0359895, &
\alpha(\MZ^2) &=& 1/128.887, &
\GF & = & 1.16637 \times 10^{-5} \GeV^{-2}, \\
\MW & = & 80.35\GeV, &
\MZ & = & 91.1867\GeV, &
\MH & = & 150\GeV, \\
\GW & = & 2.08699\ldots\GeV, & \alpha_{\mathrm{s}} &=& 0.119, \\
\Me & = & 0.51099907\MeV,  \hspace{1em} &
m_{\mu} & = & 105.658389\MeV,  \hspace{1em} &
m_{\tau} & = & 1.77705\;\GeV, \\
\Mu & = & 4.85\;\MeV, &
\Mc & = & 1.55\;\GeV, &
\Mt & = & 174.17\;\GeV, \\
\Md & = & 4.85\;\MeV, &
\Ms & = & 150\;\MeV, &
\Mb & = & 4.5\;\GeV, \\
|V_{\Pu\Pd}| & = & 0.975, &
|V_{\Pu\Ps}| & = & 0.222, \\
|V_{\Pc\Pd}| & = & 0.222, &
|V_{\Pc\Ps}| & = & 0.975, 
\end{array}
\label{eq:par}
\eeq
which is consistent with experimental data \cite{Caso:1998tx}.
Except for the CKM matrix elements, this input is identical with
the one used in the LEP2 Monte Carlo workshop report \cite{Grunewald:2000ju}
on precision calculations for LEP2.

We recall that the above set of data is overcomplete, but all the
numbers are needed for the evaluation of the ${\cal O}(\alpha)$ corrected
cross section in the different input schemes described in 
\refse{se:born}. The masses of the light quarks are only relevant for
the evaluation of the charge renormalization constant $\delta Z_e$,
which drops out in the $\alpha(\MZ)$- and $\GF$-schemes; the values
for these masses are adjusted to reproduce the hadronic contribution
to the photonic vacuum polarization \cite{Burkhardt:1995tt}.  The
value for the W-boson decay width $\GW$ is the 
${\cal O}(\alpha)$- and ${\cal O}(\alpha_{\mathrm{s}})$-corrected 
SM prediction in the $\GF$ scheme. 
Note that $\alpha_{\mathrm{s}}$ only enters the calculation of $\GW$ in the
results presented here, since we do not consider QCD corrections to
the scattering processes.  
If not stated otherwise, the presented cross sections and distributions
are calculated for a fixed W-boson width.  
We consistently take the CTEQ4L
\cite{Lai:1997mg} quark distributions for the evaluation of the
$\Pp\Pp$ and $\Pp\bar\Pp$ cross sections.
If not stated otherwise
the factorization scale $M$ is set to the W-boson mass $\MW$,
which is the standard choice for resonant W~production.
A thorough investigation of the full QCD and QED
factorization scale dependence would 
require the inclusion of the relevant QCD corrections and QED-improved 
parton densities, which is beyond the scope of this paper.
Instead we at least study the dependence of the 
${\cal O}(\alpha)$-corrected hadron cross sections on the QED
factorization scale $M$ using parton densities without QED evolution.

\subsection{Results for the parton process}

\begin{figure}
\centerline{
\setlength{\unitlength}{1cm}
\begin{picture}(16,7.6)
\put(-3,-1.1){\includegraphics{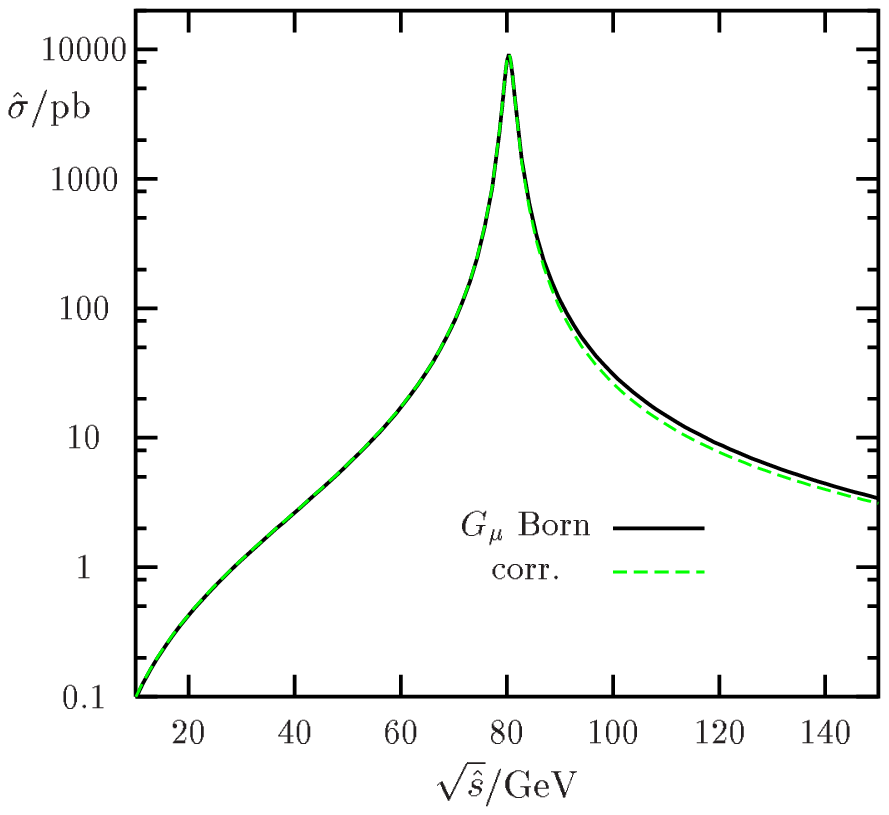}}
\put( 5,-1.1){\includegraphics{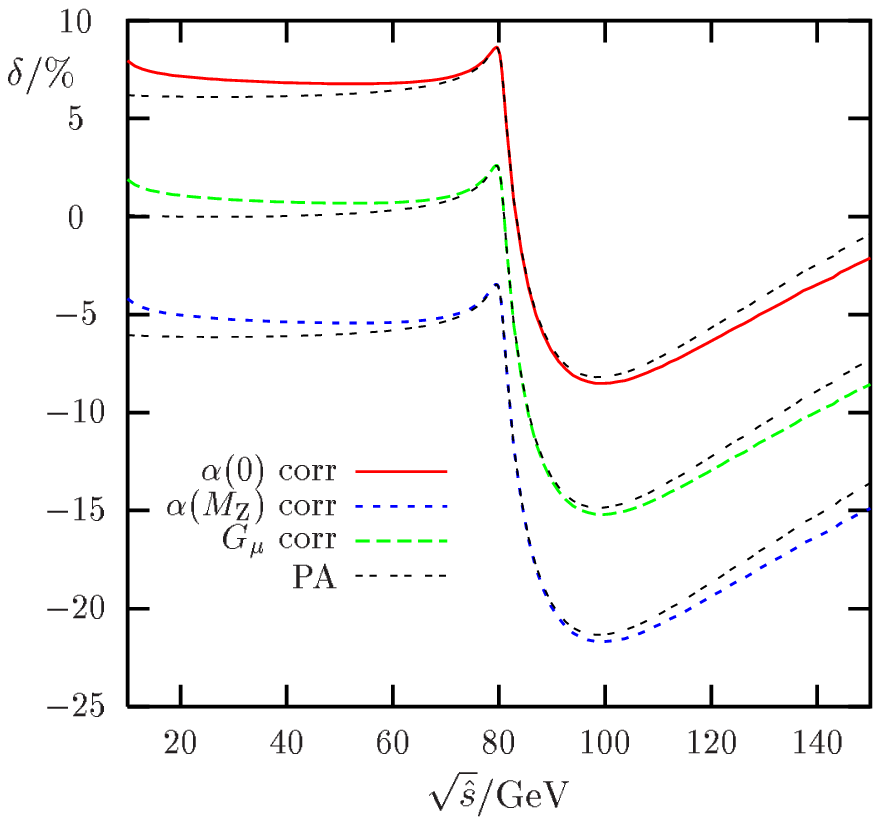}}
\put(1.2,6.5){\small $u\bar d\to\nu_l l^+(+\gamma)$}
\end{picture}
} 
\caption{Total parton cross section $\hat\sigma$ in $\GF$ 
parametrization and relative corrections $\delta$ for different
parametrizations; the respective PAs for $\delta$ are shown for
comparison.}
\label{fig:Wparton}
\end{figure}
Figure~\ref{fig:Wparton} shows the total partonic cross section
$\hat\sigma$ and the corresponding relative correction $\delta$ for
intermediate energies. Note that the total cross section (including
its correction) is the same for all final-state leptons
$l=\Pe,\mu,\tau$ in the limit of vanishing lepton masses. As expected,
the $\GF$ parametrization of the Born cross section minimizes the
correction at low energies, since the universal corrections induced by
the running of $\alpha$ and by the $\rho$ parameter are absorbed in
the lowest-order cross section.  Moreover, the naive error estimate
\refeq{eq:PAerr} for the PA turns out to be realistic. The PA
describes the correction in the resonance region within a few $0.1\%$.
\begin{table}
\centerline{
\begin{tabular}{|c||c|c|c|c|c|c|c|}
\hline
$\sqrt{\hat s}/\mathrm{GeV}$ & 40 & 80 & 120 & 200 & 500 & 1000 & 2000 
\\  \hline
$\hat\sigma_0/\mathrm{pb}$ & 2.646 & 7991.4 & 8.906 & 1.388 & 0.165 & 
0.0396 & 0.00979
\\  \hline
$\delta/\%$ & 0.7 & 2.42 & $-12.9$ & $-3.3$ & 12 & 19 & 23
\\  \hline
$\delta_\PA/\%$ & 0.0 & 2.40 & $-12.3$ & $-0.7$ & 18 & 31 & 43
\\  \hline
\end{tabular}
}
\caption{Total lowest-order parton cross section $\hat\sigma_0$ in $\GF$ 
parametrization and corresponding relative correction $\delta$, exact
and in PA.}
\label{tab:Wparton}
\end{table}
Table~\ref{tab:Wparton} contains some results on the partonic cross
section and its correction up to energies in the TeV range. Far above
resonance the PA cannot describe the exact correction anymore, since
non-resonant corrections become more and more important. 
The difference between the full ${\cal O}(\alpha)$ correction and the PA
is mainly due to (negative) Sudakov logarithms of the form
$\alpha\ln^2(\hat s/\MW^2)$, which are not contained in the PA,
but in the full ${\cal O}(\alpha)$ correction.
The large positive correction at high energies is induced by the
radiative return to the W~resonance via photon emission from the
initial-state quarks and from the W~boson, i.e.\ the reaction
proceeds as $\PW\gamma$ production with subsequent W~decay in this case.

Finally, we investigate the relation between the parametrizations of 
the W~resonance by a constant or running width. As explained in
\refse{se:born}, at tree level the transition from a running
to a constant width in the cross section is equivalent 
to the change \refeq{eq:mwdef} in $\MW$ and $\GW$.
Here we check whether, or to which accuracy, this statement 
remains valid in the presence of the $\O(\alpha)$ corrections.
\begin{figure}
\centerline{
\setlength{\unitlength}{1cm}
\begin{picture}(12,7.6)
\put(-3,-1.1){\includegraphics{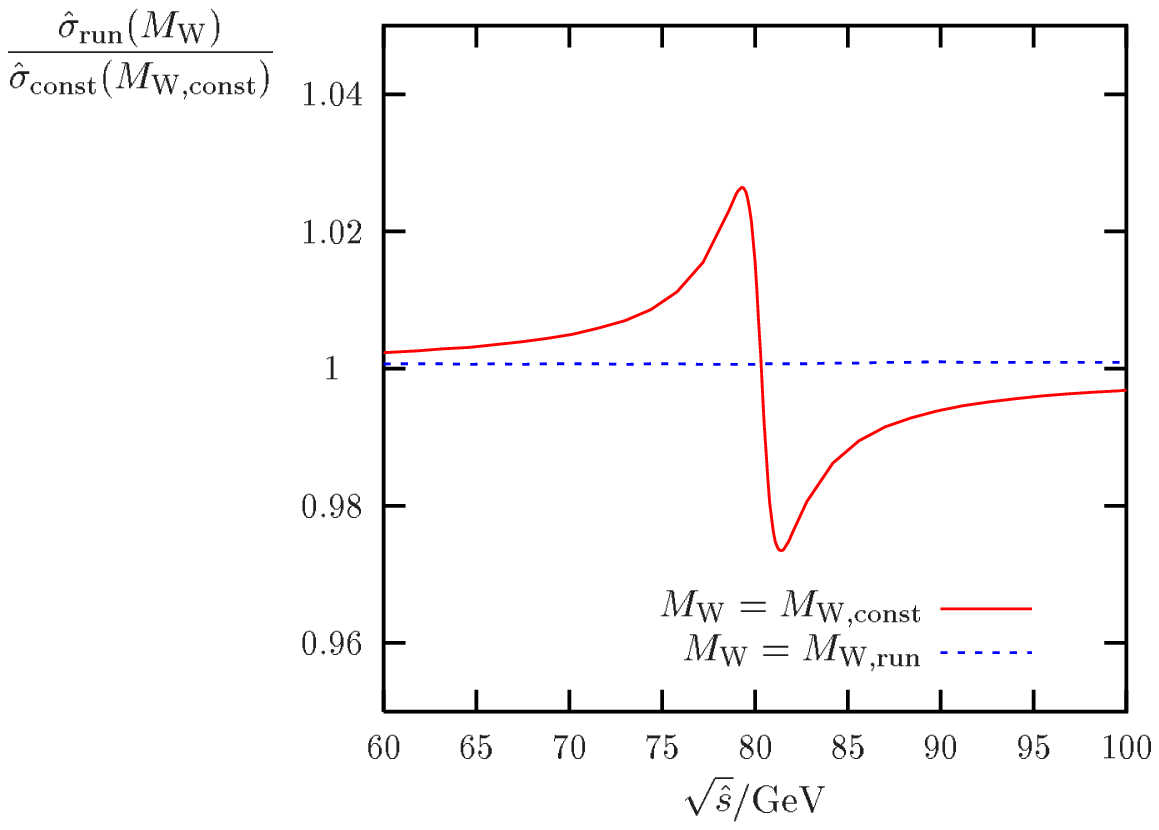}}
\put(3.7,6.5){\small $u\bar d\to\nu_\mu \mu^+(+\gamma)$}
\end{picture}
}
\caption{Ratio of $u\bar d\to\nu_\mu \mu^+(+\gamma)$ cross sections
in the $\GF$-scheme evaluated with running and constant widths
in the W~propagator.} 
\label{fig:mwdef}
\end{figure}
Figure~\ref{fig:mwdef} shows the ratio 
$\hat\sigma_{\mathrm{run}}/\hat\sigma_{\mathrm{const}}$ of the cross sections
for $u\bar d\to\nu_\mu \mu^+(+\gamma)$, evaluated with a
running width ($\hat\sigma_{\mathrm{run}}$) and with a constant width
($\hat\sigma_{\mathrm{const}}$).
Switching from one parametrization to the other without correcting
$\MW$ and $\GW$ changes the cross section at the level of a few per 
cent near the resonance, as expected. However, adjusting additionally
$\MW$ and $\GW$ according to \refeq{eq:mwdef}, absorbs this change
in the cross section up to a difference that is below 0.1\%.
This means that also in the presence of the $\O(\alpha)$ corrections
the results of a W-mass or width determination can be transformed
from one parametrization to the other via \refeq{eq:mwdef}, 
without repeating the analysis in the other parametrization.

The results of this section have been obtained with the three 
variants described in \refse{se:virt+real} for combining virtual and 
real corrections. For slicing parameters 
$2\Delta E/\sqrt{\hat s}$ and $\Delta\theta$ in the range 
$10^{-3}$--$10^{-5}$ the results agree within the integration
errors, which are significantly below the accuracy of the numbers
in \refta{tab:Wparton}.

\subsection{\boldmath{Results for $\Pp\Pp\to\PW^+\to\nu_l l^+(+\gamma)$ 
at the LHC}}

We first consider W~production at the LHC, i.e.\ we assume a pp
initial state with a CM energy of $\sqrt{s}=14\TeV$. For the
experimental identification of the process 
$\Pp\Pp\to\PW^+\to\nu_l l^+(+\gamma)$ we take the set of phase space cuts
\beq
p_{\mathrm{T},l}>25\GeV, \qquad
\dsl{p}_{\mathrm{T}}>25\GeV, \qquad
|\eta_l|<1.2,
\label{eq:lcuts}
\eeq
where $p_{\mathrm{T},l}$ and $\eta_l$ are the transverse momentum and
the rapidity of the charged lepton $l^+$, respectively, and
$\dsl{p}_{\mathrm{T}}=p_{\mathrm{T},\nu_l}$ 
is the missing transverse momentum carried away
by the neutrino.  Note that these cuts are not ``collinear-safe'' 
with respect to the lepton momentum, so
that observables in general receive corrections that involve large
lepton-mass logarithms of the form $\alpha\ln(\Ml/\MW)$. This is due
to the fact that photons within a small collinear cone around the
charged lepton momentum are not treated inclusively, i.e.\ the cuts
assume a perfect isolation of photons from the charged lepton. While
this is (more or less) achievable for muon final states, it is not
realistic for electrons.  In order to be closer to the experimental
situation for electrons, we additionally consider the following photon
recombination procedure:
\begin{enumerate}
\item
Photons with a rapidity $|\eta_\gamma| > 2.5$, which are close to 
the beams, are treated as invisible,
i.e.\ they are considered as part of the proton remnant.
\item 
If the photon survived the first step, 
and if the resolution $R_{l\gamma} = 
\sqrt{(\eta_l-\eta_\gamma)^2 + \phi_{l\gamma}^2}$
is smaller than 0.1
(with $\phi_{l\gamma}$ denoting the angle between lepton and photon
in the transverse plane),
then the photon is recombined with the charged lepton, 
i.e.\ the momentum of the photon and $l$ are added and
associated with the momentum of $l$, and the photon is discarded.
\item 
Finally, all events are discarded in which the resulting momentum of the
charged lepton does not pass the cuts given in \refeq{eq:lcuts}.
\end{enumerate}
While the electroweak corrections differ for final-state electrons and
muons without photon recombination, called ``bare'' leptons in the
following, the corrections become universal in the presence of photon
recombination, since the lepton-mass logarithms cancel in this case,
in accordance with the KLN theorem \cite{KLN}.
From the explanations at the end of \refse{se:sub} it is clear that
the described subtraction approach is not applicable without photon
recombination, i.e.\ for bare leptons, while the slicing variants
can be used with and without photon recombination.

\begin{figure}
\centerline{
\setlength{\unitlength}{1cm}
\begin{picture}(16,7.6)
\put(-3,-1.1){\includegraphics{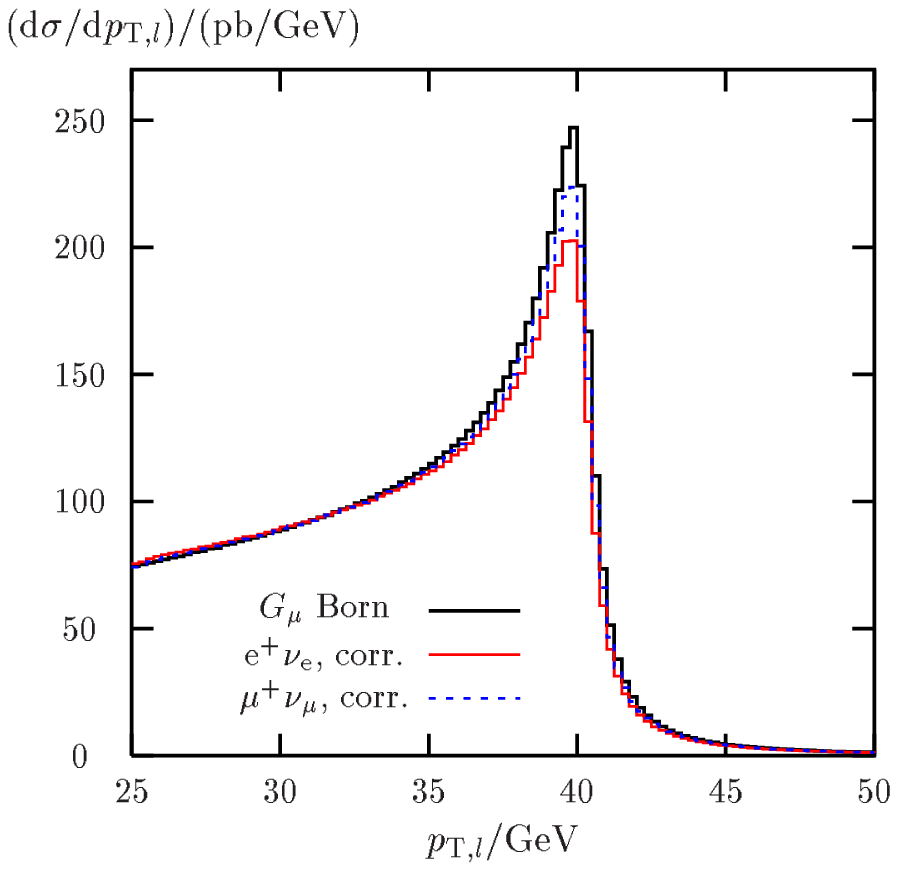}}
\put( 5,-1.1){\includegraphics{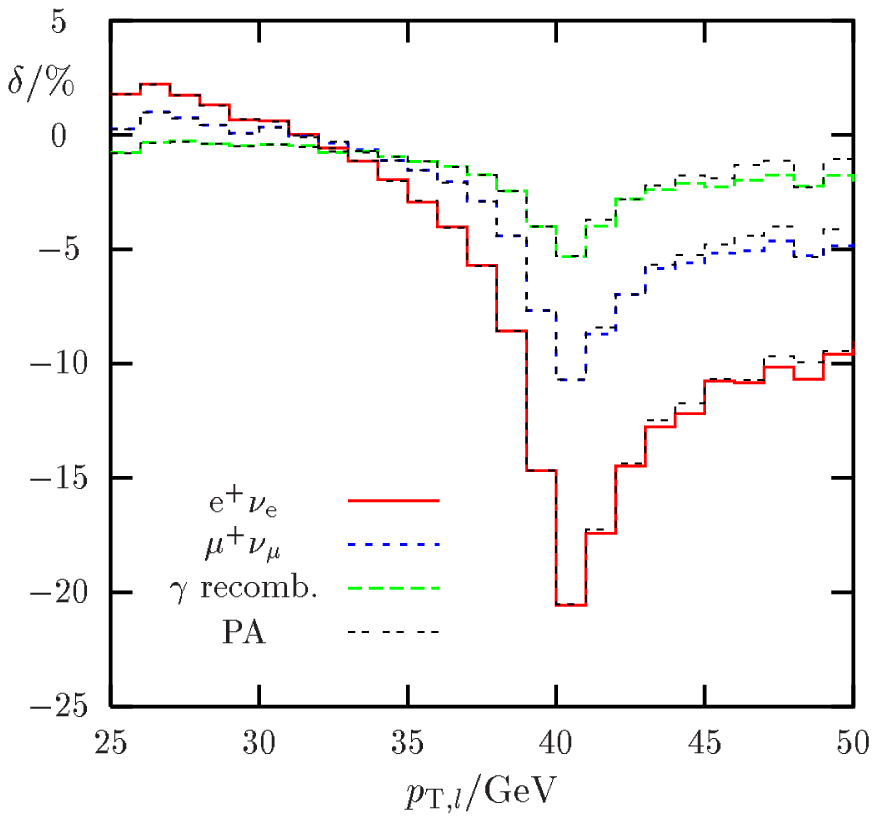}}
\put(1.3,6.6){\small $\Pp\Pp\to\nu_l l^+(+\gamma)$}
\put(1.3,6.0){\small $\sqrt{s}=14\TeV$}
\put(1.3,5.4){\small $p_{\mathrm{T},l},\dsl{p}_{\mathrm{T}}>25\GeV$}
\put(1.3,4.8){\small $|\eta_l|<1.2$}
\end{picture}
} 
\caption{Transverse-momentum distribution 
$(d\sigma/dp_{\mathrm{T},l})$ and relative corrections $\delta$.}
\label{fig:pp_ptl_lhc}
%
\vspace*{3em}
\centerline{
\setlength{\unitlength}{1cm}
\begin{picture}(16,7.6)
\put(-3,-1.1){\includegraphics{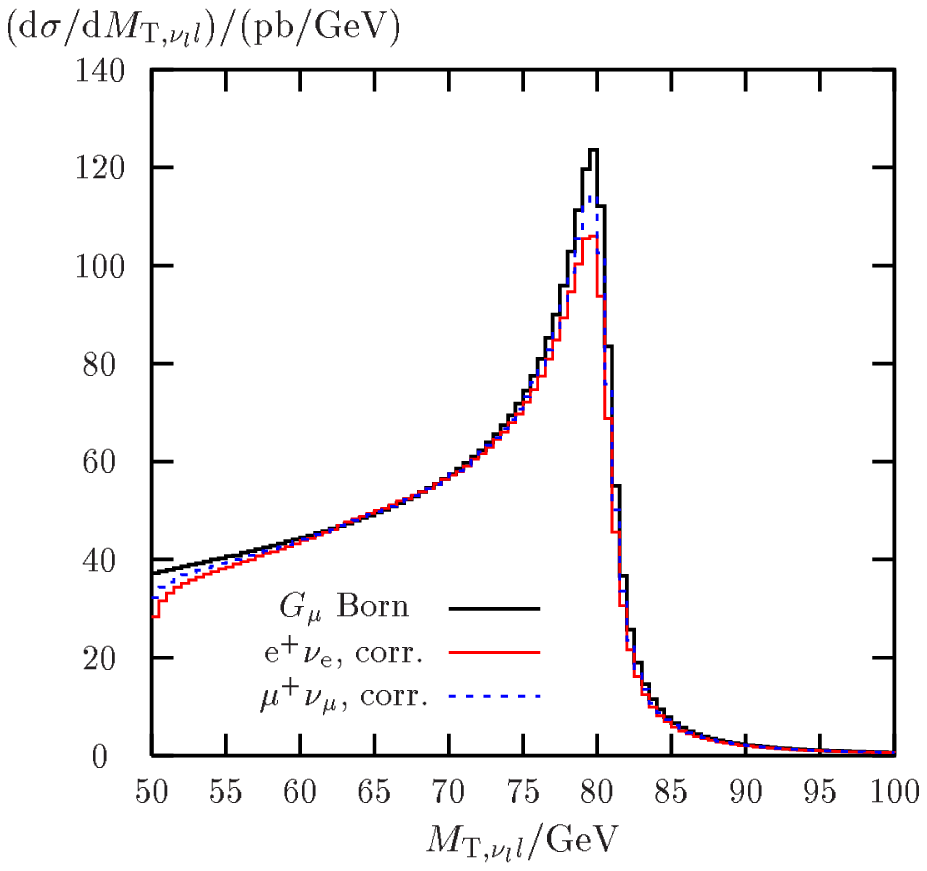}}
\put( 5,-1.1){\includegraphics{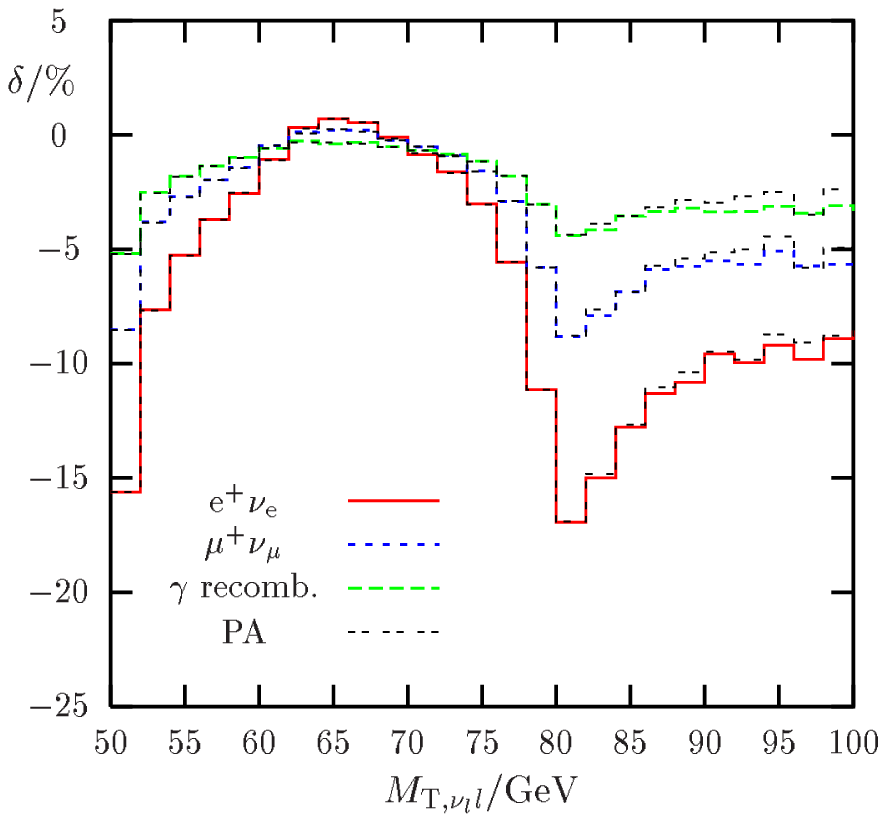}}
\put(1.3,6.6){\small $\Pp\Pp\to\nu_l l^+(+\gamma)$}
\put(1.3,6.0){\small $\sqrt{s}=14\TeV$}
\put(1.3,5.4){\small $p_{\mathrm{T},l},\dsl{p}_{\mathrm{T}}>25\GeV$}
\put(1.3,4.8){\small $|\eta_l|<1.2$}
\end{picture}
}
\caption{Transverse-invariant-mass distribution 
$(d\sigma/dM_{\mathrm{T},\nu_l l})$ and relative corrections $\delta$.}
\label{fig:pp_mtnl_lhc}
\end{figure}
Figures~\ref{fig:pp_ptl_lhc} and \ref{fig:pp_mtnl_lhc} show the
distributions in the transverse-momentum $p_{\mathrm{T},l}$ and in the
transverse-invariant-mass $M_{\mathrm{T},\nu_l l}$
in $\Pp\Pp\to\PW^+\to\nu_l l^+(+\gamma)$ for the LHC energy, together
with the corresponding relative electroweak corrections $\delta$.  
The transverse-invariant-mass is defined by $M_{\mathrm{T},\nu_l l}=
\sqrt{2p_{\mathrm{T},l}\dsl{p}_{\mathrm{T}}(1-\cos\phi_{\nu_l l})}$,
where $\phi_{\nu_l l}$ is the angle between the lepton and the
missing momentum in the transverse plane.
The distributions show the well-known kinks at $p_{\mathrm{T},l}\approx
\MW/2$ and $M_{\mathrm{T},\nu_l l}\approx\MW$, which are used in the
W-mass determination. Near these kinks the correction $\delta$ reaches
the order of 10--20\% for bare leptons, where the larger corrections
occur in the electron case, because the logarithm $\alpha\ln(\Ml/\MW)$
is larger in this case.  Since these enhanced corrections originate
from collinear final-state radiation, they are negative for higher
$p_{\mathrm{T},l}$ and redistribute events to lower transverse
momenta.  The correction $\delta$ is reduced to a few per cent after
photon recombination, which eliminates the artificial lepton-mass
logarithms.

Moreover, Figures~\ref{fig:pp_ptl_lhc} and \ref{fig:pp_mtnl_lhc}
demonstrate the reliability of the PA for transverse lepton momenta
$p_{\mathrm{T},l}\lsim \MW/2$, where resonant W~bosons dominate.
The PA curves are included in the plots
as thin double-dashed lines close to the corresponding full corrections.
Only for transverse momenta above the resonance region a systematic
difference between the PA and the full result starts to become apparent.

Transverse lepton momenta $p_{\mathrm{T},l}$ above the resonance kink
need to be considered for the
determination of the W~decay width $\GW$ and, in the high-energy tail,
for searches for new physics, such as new $\PW'$ gauge bosons.
\begin{table}
\centerline{
\begin{tabular}{|c||c|c|c|c|c|c|}
\hline
\multicolumn{7}{|l|}{$\Pp\Pp\to\nu_l l^+(+\gamma)$ at $\sqrt{s}=14\TeV$}
\\  \hline  \hline 
$p_{\mathrm{T},l}/\mathrm{GeV}$ & 
25--$\infty$ & 50--$\infty$ & 100--$\infty$ & 200--$\infty$ & 500--$\infty$ 
& 1000--$\infty$
\\  \hline 
$\sigma_0/\mathrm{pb}$ & 
1933.5(3) & 11.50(1) & 0.8198(4) & 0.1015(1) & 0.005277(1) & 0.0003019(1)
\\  \hline
$\delta_{\Pep\nu_\Pe}/\%$ &  
$-5.5(1)$ & $-9.2(1)$ & $-11.8(1)$ & $-16.5(1)$ & $-27.5(2)$ & $-39.1(1)$
\\  
$\delta_{\Pep\nu_\Pe,\PA}/\%$ &  
$-5.4(1)$ & $-8.0(1)$ & $-7.3(1)$ & $-7.8(1)$ & $-10.3(2)$ & $-13.0(1)$
\\  \hline
$\delta_{\mu^+\nu_\mu}/\%$ &  
$-2.9(1)$ & $-4.9(1)$ & $-8.5(1)$ & $-13.1(1)$ & $-23.4(1)$ & $-34.5(1)$
\\ 
$\delta_{\mu^+\nu_\mu,\PA}/\%$ &  
$-2.8(1)$ & $-3.5(1)$ & $-4.0(1)$ & $-4.4(1)$ & $-6.2(1)$ & $-8.5(1)$
\\  \hline
$\delta_{\rec}/\%$ &  
$-1.8(1)$ & $-2.7(1)$ & $-6.2(1)$ & $-10.2(1)$ & $-19.6(1)$ & $-29.6(1)$
\\
$\delta_{\rec,\PA}/\%$ &  
$-1.8(1)$ & $-1.5(1)$ & $-1.6(1)$ & $-1.6(1)$ & $-2.4(1)$ & $-3.6(1)$
\\  \hline\hline
$\Delta(2^{\pm 1}\MW)/\%$ &  
$\pm0.1$ & $\pm0.1$ & $\pm0.1$ & $\pm0.1$ & $\pm0.1$ & $\pm0.2$ 
\\
$\Delta(M_{\mathrm{T},\nu_l l})/\%$ &  
$-0.0$ & $0.0$ & $0.1$ & $0.2$ & $0.5$ & $0.8$ 
\\  \hline
\end{tabular}
}
\caption{Integrated lowest-order pp cross sections $\sigma_0$ 
for different ranges in $p_{\mathrm{T},l}$ and corresponding 
relative corrections $\delta$, exact and in PA. 
The QED scale uncertainty is illustrated by $\Delta(M)$, as
described in the text.}
\label{tab:Wpp_lhc}
\end{table}
Table~\ref{tab:Wpp_lhc} shows the high-$p_{\mathrm{T},l}$ 
contributions to the total cross section defined by different
ranges in $p_{\mathrm{T},l}$. Although the cross section rapidly decreases
for high $p_{\mathrm{T},l}$, sizeable event numbers can be expected
for $p_{\mathrm{T},l}$ values in the range 100--$1000\GeV$ at the LHC.
The results of \refta{tab:Wpp_lhc} reveal that the PA is not 
applicable for very large $p_{\mathrm{T},l}$, where the W~boson is far 
off shell. This fact was also to be expected from the results at the
parton level discussed in the previous section.
As mentioned previously, the breakdown of the PA is mainly due to the 
missing Sudakov logarithms, which are independent of the lepton species.
This explains why the PA breaks down both for electrons and muons in the
same way, with or without photon recombination.
Moreover, these large negative corrections of up to $\sim 30\%$
for $p_{\mathrm{T},l}$ values between 500--$1000\GeV$ do not
become small after photon recombination. 
Note also that the large positive correction to the total
parton cross section at high scattering energies
(see \refta{tab:Wparton}), which is mainly due to resonant 
$W\gamma$ production in the forward and backward directions,
is widely suppressed at the hadron level. The reason for this
suppression is the cut on the transverse lepton momentum which
effectively demands large W~momenta.

Table~\ref{tab:Wpp_lhc} additionally shows the uncertainty
of the ${\cal O}(\alpha)$-corrected cross sections that is due
to the dependence on the QED factorization scale $M$, as described
in \refse{se:ppcs}. The quantity
\beq
\Delta(M) = \delta(M)-\delta
\eeq
quantifies the difference of the relative correction $\delta(M)$,
evaluated for the scale $M$, to its value $\delta$
for the default choice $M=\MW$.
This difference $\Delta(M)$ applies to each row in \refta{tab:Wpp_lhc},
since the $M$-dependent part is the same for each version of
$\delta_{\dots}$ shown in the table.
We have chosen the values $M=2^{\pm 1}\MW$ and
$M=M_{\mathrm{T},\nu_l l}$
for illustration, the former representing a simple rescaling of our
default choice, the latter a scale choice that might be more
appropriate for large transverse momenta.
Although the parton densities have not been corrected for the QED
factorization (see \refse{se:ppcs}), the QED
scale dependence turns out to be negligible
compared to the other theoretical uncertainties and the expected
experimental accuracy.
Using QED-improved parton densities the QED factorization scale
dependence should be reduced further.

The results without photon recombination have been obtained with
the two slicing variants of \refse{se:virt+real} with the same values
for the cut parameters as at the parton level. The results
with photon recombination have also been checked with the subtraction
approach. For the distributions shown in \reffis{fig:pp_ptl_lhc} and
\ref{fig:pp_mtnl_lhc}, the results from the different methods agree 
at the level of the line width in the plots.%
\footnote{Of course, the statistical error of the histogram bins
is somewhat larger than the line width, at least for large transverse 
momenta and invariant masses. This means that the differences between
the subtraction and slicing variants are even smaller than the statistical
error. However, this simply results from the fact that in all cases the
same set of random numbers was used, which introduces correlations
among the statistical errors in the contributions that are not affected
by subtraction or slicing.}
The results of \refta{tab:Wpp_lhc} have been obtained with the various
methods with an agreement within 1--2 standard deviations, apart from
a few exceptions due to fluctuations (signalled by a large 
$\chi^2/\mathrm{d.o.f.}$). Apart from those, all numbers without
photon recombination result from 
the ``effective collinear factor'' approach,
while the ones with photon recombination are obtained with subtraction.

\subsection{\boldmath{Results for $\Pp\bar\Pp\to\PW^+\to\nu_l l^+(+\gamma)$ at 
the Tevatron}}

Now we consider W~production at the Tevatron, i.e.\ at a $\Pp\bar\Pp$
collider with a CM energy of $\sqrt{s}=2\TeV$. 
We again use the phase space of \refeq{eq:lcuts} and the photon
recombination procedure%
\footnote{More realistic electron and muon identification procedures
for the Tevatron have been used in \citere{Baur:1999kt}, where
electroweak corrections to the transverse momentum and invariant-mass
distributions have been evaluated in another variant of the pole
approximation.}
of the previous section.

\begin{figure}
\centerline{
\setlength{\unitlength}{1cm}
\begin{picture}(16,7.6)
\put(-3,-1.1){\includegraphics{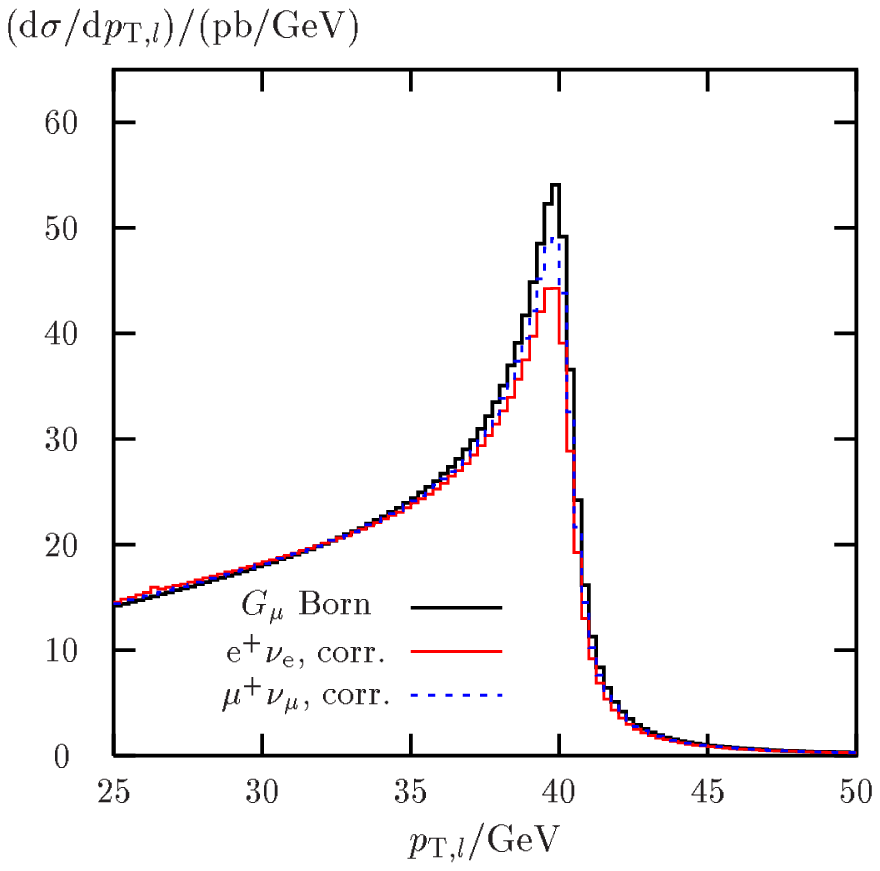}}
\put( 5,-1.1){\includegraphics{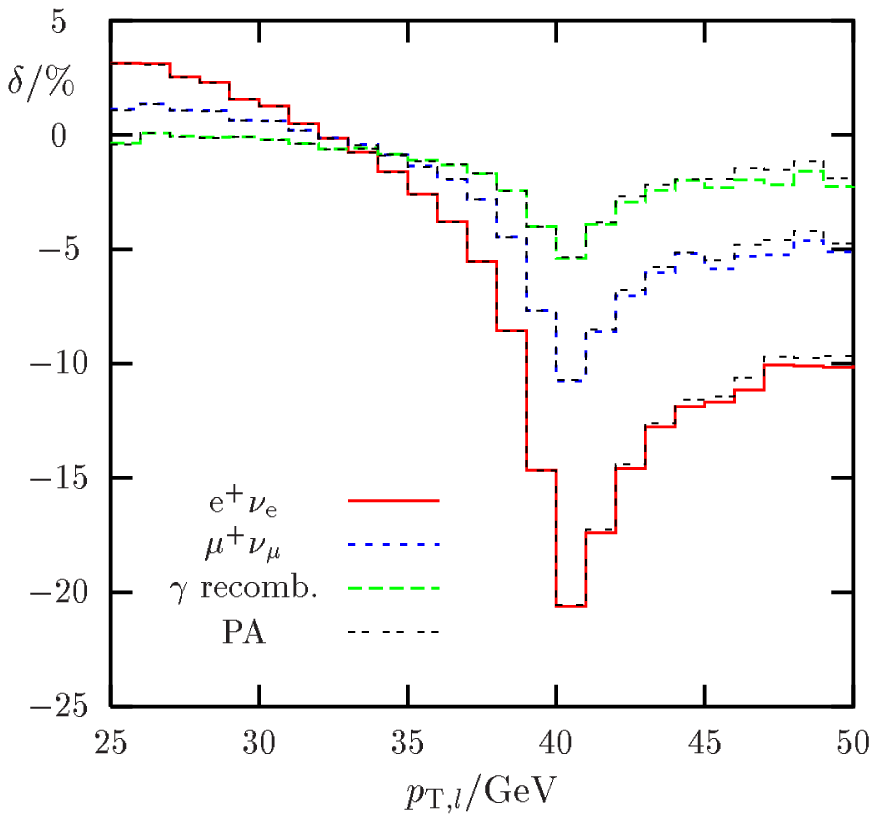}}
\put(1.3,6.6){\small $\Pp\bar\Pp\to\nu_l l^+(+\gamma)$}
\put(1.3,6.0){\small $\sqrt{s}=2\TeV$}
\put(1.3,5.4){\small $p_{\mathrm{T},l},\dsl{p}_{\mathrm{T}}>25\GeV$}
\put(1.3,4.8){\small $|\eta_l|<1.2$}
\end{picture}
} 
\caption{Transverse-momentum distribution 
$(d\sigma/dp_{\mathrm{T},l})$ and relative corrections $\delta$.}
\label{fig:pp_ptl_tev}
%
\vspace*{3em}
\centerline{
\setlength{\unitlength}{1cm}
\begin{picture}(16,7.6)
\put(-3,-1.1){\includegraphics{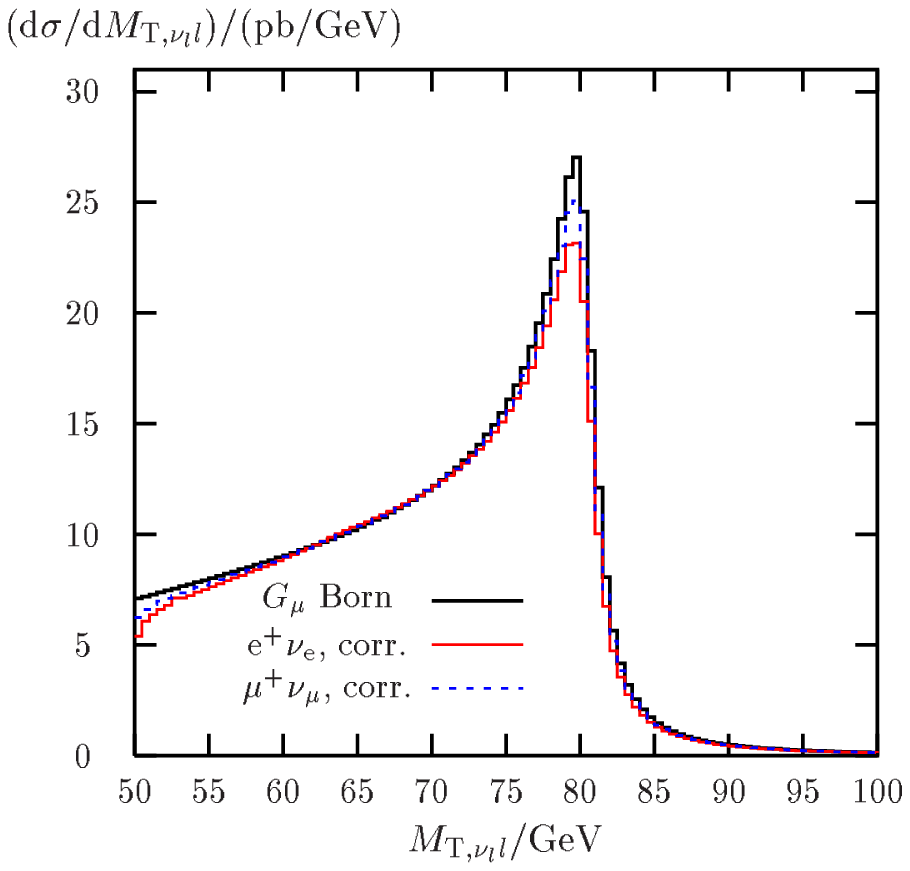}}
\put( 5,-1.1){\includegraphics{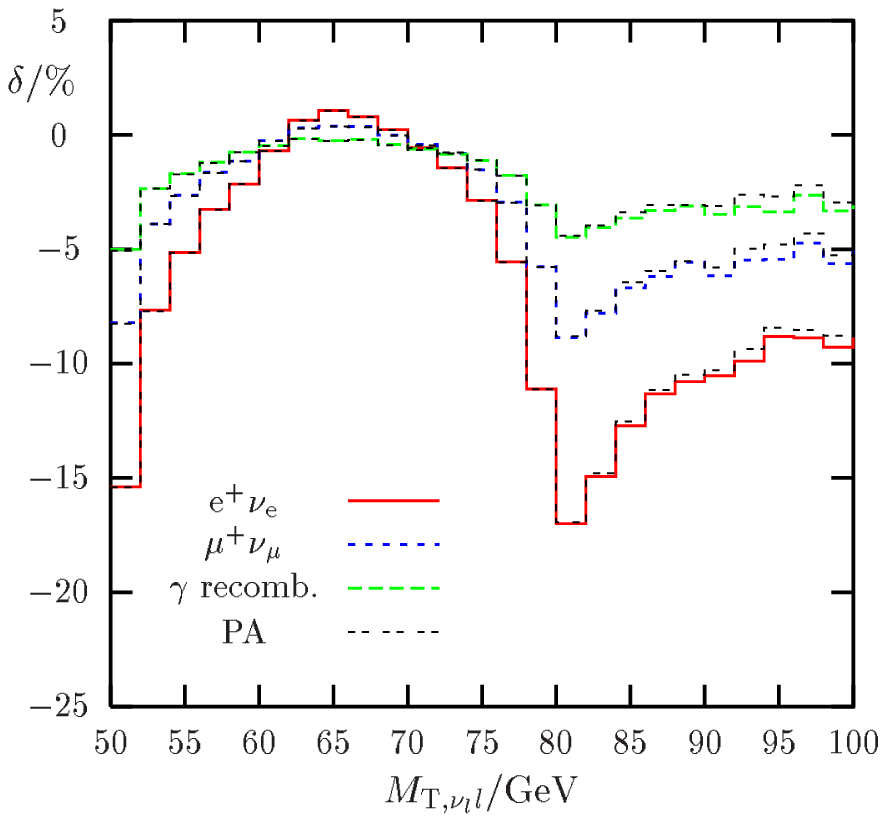}}
\put(1.3,6.6){\small $\Pp\bar\Pp\to\nu_l l^+(+\gamma)$}
\put(1.3,6.0){\small $\sqrt{s}=2\TeV$}
\put(1.3,5.4){\small $p_{\mathrm{T},l},\dsl{p}_{\mathrm{T}}>25\GeV$}
\put(1.3,4.8){\small $|\eta_l|<1.2$}
\end{picture}
}
\caption{Transverse-invariant-mass distribution 
$(d\sigma/dM_{\mathrm{T},\nu_l l})$ and relative corrections $\delta$.}
\label{fig:pp_mtnl_tev}
\end{figure}
The corrections to the transverse-momentum and transverse-invariant-mass 
distributions at intermediate transverse momenta,
displayed in \reffis{fig:pp_ptl_tev} and \ref{fig:pp_mtnl_tev},
are of the same size as the ones discussed for the LHC in the 
previous section. They also show the same qualitative features under
the changes in the lepton mass and after photon recombination.
Moreover, the results for the PA are again not distinguishable from the
full correction in the plots as long as resonant W~production is
possible; only for transverse momenta $p_{\mathrm{T},l}$ much larger
than $\MW/2$ differences become visible.

\begin{table}
\centerline{
\begin{tabular}{|c||c|c|c|c|c|c|}
\hline
\multicolumn{7}{|l|}{$\Pp\bar\Pp\to\nu_l l^+(+\gamma)$ at $\sqrt{s}=2\TeV$}
\\  \hline  \hline 
$p_{\mathrm{T},l}/\mathrm{GeV}$ & 
25--$\infty$ & 50--$\infty$ & 75--$\infty$ & 100--$\infty$ & 200--$\infty$ 
& 300--$\infty$
\\  \hline
$\sigma_0/\mathrm{pb}$ & $407.03(5)$ & $2.481(1)$ & 
$0.3991(1)$ & $0.1305(1)$ & $0.006020(2)$ & $0.0004821(1)$
\\  \hline
$\delta_{\Pep\nu_\Pe}/\%$ &  
$-5.4(1)$ & $-9.7(1)$ & $-11.2(1)$ & $-12.9(1)$ & $-18.8(1)$ & $-23.8(1)$
\\  
$\delta_{\Pep\nu_\Pe,\PA}/\%$ &  
$-5.3(1)$ & $-8.6(1)$ & $-8.7(1)$ & $-9.2(1)$ & $-11.7(1)$ & $-14.1(1)$
\\  \hline
$\delta_{\mu^+\nu_\mu}/\%$ &  
$-2.9(1)$ & $-5.3(1)$ & $-7.3(1)$ & $-9.0(1)$ & $-14.2(1)$ & $-18.5(1)$
\\ 
$\delta_{\mu^+\nu_\mu,\PA}/\%$ &  
$-2.8(1)$ & $-4.2(1)$ & $-4.8(1)$ & $-5.2(1)$ & $-7.1(1)$ & $-8.8(1)$
\\  \hline
$\delta_{\rec}/\%$ &  
$-1.8(1)$ & $-2.7(1)$ & $-4.8(1)$ & $-6.3(1)$ & $-10.4(1)$ & $-13.6(1)$
\\
$\delta_{\rec,\PA}/\%$ &  
$-1.7(1)$ & $-1.6(1)$ & $-2.3(1)$ & $-2.5(1)$ & $-3.3(1)$ & $-3.9(1)$
\\  \hline\hline
$\Delta(2^{\pm 1}\MW)/\%$ &  
$\pm0.1$ & $\pm0.1$ & $\pm0.1$ & $\pm0.1$ & $\pm0.2$ & $\pm0.2$ 
\\
$\Delta(M_{\mathrm{T},\nu_l l})/\%$ &  
$-0.0$ & $0.1$ & $0.1$ & $0.2$ & $0.5$ & $0.7$ 
\\  \hline
\end{tabular}
}
\caption{Integrated lowest-order $\Pp\bar\Pp$ cross sections $\sigma_0$ 
for different ranges in $p_{\mathrm{T},l}$ and corresponding 
relative corrections $\delta$, exact and in PA.
The QED scale uncertainty is illustrated by $\Delta(M)$, as
described in thet text.}
\label{tab:Wpp_tev}
\end{table}
The corrections to the high-transverse momentum tail are illustrated in
\refta{tab:Wpp_tev}. As expected, the PA again becomes worse with 
increasing $p_{\mathrm{T},l}$, but owing to the lower luminosity and the
smaller cross section at the Tevatron in comparison to the LHC it will be 
extremely hard to see any effect of the enhanced electroweak corrections
at high $p_{\mathrm{T},l}$.
Concerning the QED factorization scale dependence, which is again 
quantified by $\Delta(M)$, the same remarks apply that have already 
been made for the LHC case.
The QED scale uncertainty turns out to be negligible.

\section{Conclusions}
\label{se:concl}

We have calculated the complete set of electroweak ${\cal O}(\alpha)$
corrections to the Drell--Yan-like W-boson production at hadron
colliders.  Particular attention has been paid to issues of gauge
invariance and the instability of the W~bosons.  All relevant formulae
are listed in a form that facilitates their implementation in computer
codes.  Besides results for the full correction for off-shell
W~bosons, we have also presented an expansion of the virtual
correction about the W-resonance pole, which is considerably simpler
than the full result.

Numerical results have been discussed at the parton level and for
hadronic collisions at the LHC and at the Tevatron. The electroweak
corrections significantly influence the transverse momentum and
invariant-mass distributions of the decay leptons that are used in the
determination of the W-boson mass. The pole approximation yields a
good description of the corrections to these observables.  This result
justifies, in particular, the present practice at the Tevatron, where such
an approximation is used in the W-mass measurement.  However, in the
domains of non-resonant W-boson production
in these distributions, which are relevant 
for the measurement of the W-boson width or 
for the search of new-physics
effects at the LHC, this approximation fails, rendering the complete
correction important; the ${\cal O}(\alpha)$ corrections reduce the
signal by several $10\%$ for transverse lepton momenta with
$p_{\mathrm{T},l} \gsim 100\GeV$. To further improve the analysis,
in particular in
the large-$p_{\mathrm{T}}$ domain, more theoretical studies of
electroweak higher-order effects as well as realistic experimental
simulations are desirable.

\subsection*{Note added}

While this paper was completed, another calculation of the electroweak
$\O(\alpha)$ corrections to W production, including non-resonant
contributions, has been presented in \citere{Zykunov:2001mn}. The numerical
results mainly focus on observables measured at RHIC, but also include the
transverse lepton momentum distribution at the Tevatron. However, no mass
factorization has been performed in \citere{Zykunov:2001mn}, so that the
corrections depend very sensitively on the light quark masses. 
A direct comparison with the numerical results presented in our paper 
is thus not possible.

\section*{Acknowledgement}

We would like to thank U.~Baur, M.~Dittmar, W.~Hollik, H.~Spiesberger, and 
D.~Wackeroth for helpful discussions.  This work has been
supported in part by the European Union under contract
HPRN-CT-2000-00149.

\appendix
\section*{Appendix}

\section{Vertex and box corrections}
\label{app:virtRCs}

\subsection{Form factor for the \boldmath{$Wff'$} vertex}
\label{app:vert}

The form factor $F_{Wff'}(\hat s)$ used in \refeq{eq:dvert} for the
vertex corrections is explicitly given by
\beqar
F_{Wff'}(\hat s) &=& \frac{\alpha}{16\pi} \biggl\{
\frac{2(1-2\sw^2-4Q_f Q_{f'}\sw^2)}{\cw^2\sw^2}
+\frac{4}{\sw^2} \biggl(2 +\frac{\MW^2}{\hat s}\biggr) B_0(0,0,\MW)
\nn\\ && \quad {}
+\frac{2}{\cw^2\sw^2} (1-2 \sw^2+2 Q_f^2 \sw^4+2 Q_{f'}^2 \sw^4)
\biggl(2+\frac{\MZ^2}{\hat s}\biggr) B_0(0,0,\MZ)
\nn\\ && \quad {}
-4\biggl(1+\frac{\MW^2}{\hat s}\biggr) B_0(\hat s,0,\MW)
-\frac{4\cw^2}{\sw^2} \biggl(1+\frac{\MW^2+\MZ^2}{\hat s}\biggr)
  B_0(\hat s,\MW,\MZ)
\nn\\ && \quad {}
+\frac{1}{\cw^2\sw^2} \biggl[
2\frac{\MZ^2}{\hat s} (Q_f-Q_{f'}-2Q_f \sw^2)(Q_f-Q_{f'}+2Q_{f'}\sw^2)
\nn\\ && \qquad {}
+3(1-2\sw^2-4Q_f Q_{f'}\sw^2) \biggr] B_0(\hat s,0,0)
\nn\\ && \quad {}
+8Q_f^2 B_0(m_f^2,0,m_f) + 8Q_{f'}^2 B_0(m_{f'}^2,0,m_{f'})
\nn\\ && \quad {}
-8Q_f Q_{f'}\hat s C_0(m_f^2,m_{f'}^2,\hat s,m_f,m_\gamma,m_{f'})
\nn\\ && \quad {}
+8Q_f (Q_f-Q_{f'}) \MW^2 C_0(m_f^2,0,\hat s,0,m_f,\MW)        
\nn\\ && \quad {}
-8Q_{f'} (Q_f-Q_{f'}) \MW^2 C_0(m_{f'}^2,0,\hat s,0,m_{f'},\MW)        
\nn\\ && \quad {}
+\frac{8\cw^2}{\sw^2}\biggl(\MW^2+\MZ^2+\frac{\MW^2\MZ^2}{\hat s}\biggr)
C_0(0,0,\hat s,\MW,0,\MZ)
\nn\\ && \quad {}
+\frac{2}{\cw^2\sw^2} (Q_f-Q_{f'}-2Q_f\sw^2)(Q_f-Q_{f'}+2Q_{f'}\sw^2)
\nn\\ && \qquad {}
\times \biggl(1+\frac{\MZ^2}{\hat s}\biggr)^2
\hat s C_0(0,0,\hat s,0,\MZ,0)
\biggr\},
\eeqar
where the weak isospins $I^3_{\PW,f}$ and $I^3_{\PW,f'}$ of the fermions
are implicitly taken to be 
$Q_f-Q_{f'}=2I^3_{\PW,f}=-2I^3_{\PW,f'}$. Here and in the following the
scalar integrals $B_0$, $C_0$, and $D_0$ depend on their arguments as
follows,
\beqar
&& B_0(p_1^2,m_0,m_1)=
\frac{(2\pi\mu)^{4-D}}{\ri\pi^2}\int\rd^D q
\frac{1}{[q^2-m_0^2+\ri\epsilon][(q+p_1)^2-m_1^2+\ri\epsilon]},
\nn\\[.5em]
&& C_0(p_1^2,(p_2-p_1)^2,p_2^2,m_0,m_1,m_2)=
\frac{1}{\ri\pi^2}\int d^4 q 
\nn \\[.2em]
&& \hspace{1em}
\times\frac{1}{[q^2-m_0^2+\ri\epsilon]
[(q+p_1)^2-m_1^2+\ri\epsilon] [(q+p_2)^2-m_2^2+\ri\epsilon]}, 
\nn \\[.5em]
&& 
D_0(p_1^2,(p_2-p_1)^2,(p_3-p_2)^2,p_3^2,
p_2^2,(p_3-p_1)^2,m_0,m_1,m_2,m_3)=
\frac{1}{\ri\pi^{2}}\int d^{4}q 
\nn \\[.2em]
&& \hspace{1em}
\times\frac{1}{[q^2-m_0^2+\ri\epsilon] [(q+p_1)^2-m_1^2+\ri\epsilon]
[(q+p_2)^2-m_2^2+\ri\epsilon] [(q+p_3)^2-m_3^2+\ri\epsilon]}.
\hspace{3em}
\eeqar
Explicit representations for the regular integrals can, e.g., be found
in \citere{Denner:1993kt,'tHooft:1979xw}. 
The scalar integrals that involve mass-singular logarithms such as
$\ln(m_f)$ or on-shell singularities such as 
$\ln(\hat s-\MW^2+\ri\epsilon)$ are given by
\beqar
B_0(m_1^2,0,m_1) &=& \Delta + 2 + \ln\biggl(\frac{\mu^2}{m_1^2}\biggr),
\nn\\ 
B_0(\hat s,0,\MW) &=& \Delta + 2 + \ln\biggl(\frac{\mu^2}{\MW^2}\biggr)
+\biggl(\frac{\MW^2}{\hat s}-1\biggr)
\ln\biggl(1-\frac{\hat s}{\MW^2}-\ri\epsilon\biggr),
\nn\\
C_0(m_1^2,m_2^2,r,m_1,m_\gamma,m_2) &=&
\frac{1}{r}\biggl\{
\ln\biggl(-\frac{r}{m_1 m_2}-\ri\epsilon\biggr)
\ln\biggl(-\frac{r}{m_\gamma^2}-\ri\epsilon\biggr)
-\frac{1}{4}\ln\biggl(-\frac{r}{m_1^2}-\ri\epsilon\biggr)^2
\nn\\ && \quad {}
-\frac{1}{4}\ln\biggl(-\frac{r}{m_2^2}-\ri\epsilon\biggr)^2
-\frac{\pi^2}{6} \biggr\},
\nn\\ 
C_0(m_1^2,0,\hat s,0,m_1,\MW) &=&
\frac{1}{\hat s}\biggl\{
\ln\biggl(\frac{\hat s}{m_1^2}\biggr)
\ln\biggl(1-\frac{\hat s}{\MW^2}-\ri\epsilon\biggr)
\nn\\ && \quad {}
+\Li\biggl(1-\frac{\hat s}{\MW^2}-\ri\epsilon\biggr)-\frac{\pi^2}{6}
\biggr\},
\nn\\ 
C_0(m_1^2,0,\MW^2,m_\gamma,m_1,\MW) &=&
\frac{1}{\MW^2} \ln\biggl(\frac{\MW}{m_1}\biggr)
\ln\biggl(\frac{m_\gamma^2}{m_1\MW} \biggr),
\eeqar
where $m_\gamma^2 \ll m_{1,2}^2 \ll \hat s,|r|,\MW^2$. The quantity
$\Delta=\mbox{$2/(4-D)$}-\gamma_{\mathrm{E}}+\ln(4\pi)$ is the standard UV
divergence for $D\to 4$ space-time dimensions, and $\mu$ is the 
(arbitrary) reference mass scale of dimensional regularization.

\subsection{Box correction}
\label{app:box}

In order to reduce the Dirac structure of the box diagrams to the
structure 
\beq
c_0 = \left[ \bar v_d\gamma^\mu\omega_-u_u\right] \,
\left[ \bar u_{\nu_l}\gamma_\mu\omega_-v_l\right],
\eeq
which appears in the lowest-order matrix element $\M_0$, we used the
identities
\beqar
\left[ \bar v_d \dsl k_n \omega_-u_u\right] \,
\left[ \bar u_{\nu_l} \dsl p_u \omega_-v_l\right]
&=& -\hat t c_0/2,
\nn\\
\left[ \bar v_d \gamma^\mu \gamma^\nu \gamma^\rho \omega_-u_u\right] \,
\left[ \bar u_{\nu_l} \gamma_\mu \gamma_\nu \gamma_\rho \omega_-v_l\right]
&=& 16 c_0,
\nn\\
\left[ \bar v_d \gamma^\mu \gamma^\nu \dsl k_n \omega_-u_u\right] \,
\left[ \bar u_{\nu_l} \gamma_\mu \gamma_\nu \dsl p_u \omega_-v_l\right]
&=& -4\hat t c_0,
\eeqar
which are valid in four space-time dimensions.

\newcommand{\GNM}{g_{\nu_l}^-}
\newcommand{\GLM}{g_l^-}
\newcommand{\GUM}{g_u^-}
\newcommand{\GDM}{g_d^-}
The box correction factor $\delta_{\mathrm{box}}(\hat s,\hat t)$
introduced in \refeq{eq:dvirt} is explicitly given by
\beqar
\lefteqn{
\delta_{\mathrm{box}}(\hat s,\hat t) =
\frac{\alpha}{4\pi}(\hat s-\MW^2) \biggl\{	}
\nn\\* && \quad
(\GDM\GLM+\GNM\GUM) \biggl[ \;
\frac{2}{\hat u} \Big(B_0(\hat t,0,0)-B_0(\hat s,\MW,\MZ)\Big)
  +4 C_0(0,0,\hat s,\MW,0,\MZ)
\nn\\ && \qquad {}
  -\frac{\MW^2+\MZ^2+\hat s+2\hat t}{\hat u^2}
    \Big(\hat t C_0(0,0,\hat t,0,\MW,0) +\hat t C_0(0,0,\hat t,0,\MZ,0)
\nn\\ && \qquad\quad {}
    +2 \hat s C_0(0,0,\hat s,\MW,0,\MZ) \Big)
\nn\\ && \qquad {}
  -\frac{\hat t(\MW^2+\MZ^2+\hat s+2\hat t)^2
    +2(\MW^2+\hat t)(\MZ^2+\hat t) \hat u}{\hat u^2}
    D_0(0,0,0,0,\hat t,\hat s,0,\MW,0,\MZ)
    \;\biggr]
\nn\\ && \quad {}
-2(\GDM\GNM+\GLM\GUM) \Big[ \, 
2 C_0(0,0,\hat s,\MW,0,\MZ)
-\hat u D_0(0,0,0,0,\hat u,\hat s,0,\MW,0,\MZ)		\, \Big]
\nn\\ && \quad {}
+Q_d Q_l \biggl[ \;
  \frac{2}{\hat u} \Big(B_0(\hat t,0,0)-B_0(\hat s,0,\MW)\Big)
    +2C_0(m_d^2,0,\hat s,0,m_d,\MW)
\nn\\ && \qquad {}
  +2C_0(\Ml^2,0,\hat s,0,\Ml,\MW)
  -\frac{\MW^2+\hat s+2\hat t}{\hat u^2}
    \Big(\hat s C_0(m_d^2,0,\hat s,0,m_d,\MW)
\nn\\ && \qquad\quad {}
    +\hat s C_0(\Ml^2,0,\hat s,0,\Ml,\MW)
    +\hat t C_0(0,0,\hat t,0,\MW,0)
    +\hat t C_0(m_d^2,\Ml^2,\hat t,m_d,m_\gamma,\Ml) \Big)
\nn\\ && \qquad {}
  -\hat t\biggl( 1+\frac{(\MW^2+\hat t)^2}{\hat u^2} \biggr)  
D_0(m_d^2,\Ml^2,0,0,\hat t,\hat s,m_d,m_\gamma,\Ml,\MW)     \; \biggr]
\nn\\ && \quad {}
-2 Q_u Q_l \biggl[ \;
  C_0(\Ml^2,0,\hat s,0,\Ml,\MW)
  +C_0(m_u^2,0,\hat s,0,m_u,\MW)
\nn\\ && \qquad {}
-\hat u D_0(m_u^2,\Ml^2,0,0,\hat u,\hat s,m_u,m_\gamma,\Ml,\MW)     \; \biggr]
\;\;\biggr\},
\eeqar
where $\hat u=-\hat s-\hat t$, and $g_f^-$ are the left-handed couplings 
of fermion $f$ to the Z~boson,
\beq
g_f^- = \frac{I^3_{\PW,f}}{\cw\sw} - \frac{\sw}{\cw} Q_f.
\eeq
The singular $D_0$ function is given by
\beqar
\lefteqn{
D_0(m_1^2,m_2^2,0,0,r,\hat s,m_1,m_\gamma,m_2,\MW) = 
\frac{1}{r(\MW^2-\hat s)} \biggl\{	
 \ln^2\biggl(\frac{m_1}{\MW}\biggr)
+\ln^2\biggl(\frac{m_2}{\MW}\biggr)
} \hspace*{4em} &&
\nn\\ &&
+2\ln\biggl(-\frac{m_1 m_2}{r}\biggr) 
\ln\biggl(\frac{\MW^2-\hat s}{m_\gamma\MW}-\ri\epsilon\biggr)
+\frac{\pi^2}{3}
+\Li\biggl(1+\frac{\MW^2}{r}\biggr) \biggr\},
\hspace*{2em}
\eeqar
with $r=\hat t,\hat u <0$.


\begin{thebibliography}{99}
\frenchspacing
\newcommand{\epj}[3]{{\sl Eur. Phys. J.} {\bf #1} (19#2) #3}
\newcommand{\zp}[3]{{\sl Z. Phys.} {\bf #1} (19#2) #3}
\newcommand{\np}[3]{{\sl Nucl. Phys.} {\bf #1} (19#2) #3}
\newcommand{\phm}[3]{{\sl Phil. Mag.} {\bf #1} (19#2) #3}
\newcommand{\pl}[3]{{\sl Phys. Lett.} {\bf #1} (19#2) #3}
\newcommand{\pr}[3]{{\sl Phys. Rev.} {\bf #1} (19#2) #3}
\newcommand{\prep}[3]{{\sl Phys.\ Rep.} {\bf #1} (19#2) #3}
\newcommand{\prl}[3]{{\sl Phys. Rev. Lett.} {\bf #1} (19#2) #3}
\newcommand{\prs}[3]{{\sl Proc. Roy. Soc.} {\bf #1} (19#2) #3}
\newcommand{\fp}[3]{{\sl Fortschr. Phys.} {\bf #1} (19#2) #3}
\newcommand{\cpc}[3]{{\sl Comput. Phys. Commun.} {\bf #1} (19#2) #3}
\newcommand{\ijmp}[3]{{\sl Int. J. Mod. Phys.} {\bf #1} (19#2) #3}
\newcommand{\nim}[3]{{\sl Nucl. Instr. Meth.} {\bf #1} (19#2) #3}
\newcommand{\nc}[3]{{\sl Nuovo Cimento} {\bf #1} (19#2) #3}
\newcommand{\vj}[4]{{\sl #1} {\bf #2} (19#3) #4}
\newcommand{\jcp}[3]{{\sl J. Comp. Phys.} {\bf #1} (19#2) #3}

\bibitem{ai99}
A.~Airapetian {\it et al.} (ATLAS Collaboration), 
``ATLAS Detector and Physics Performance: Technical Design Report, 2'',
CERN-LHCC-99-15, ATLAS-TDR-15.

\bibitem{lep2repWmass}
Z. Kunszt {\it et al.}, 
in {\sl Physics at LEP2}, eds.\ G.~Altarelli, T.~Sj\"o\-strand and
F.~Zwirner (CERN 96-01, Geneva, 1996), Vol.~1, p.~151
[hep-ph/9602352].

\bibitem{ai96}
H.~Aihara {\it et al.}, 
``Future Electroweak Physics at the Fermilab Tevatron: Report of the
TEV-2000 Study Group'', 
FERMILAB-Pub-96/082.

\bibitem{Accomando:1998wt}
E.~Accomando {\it et al.}  [ECFA/DESY LC Physics Working Group
Collaboration],
Phys.\ Rept.\  {\bf 299} (1998) 1
[hep-ph/9705442]; \\
J.~A.~Aguilar-Saavedra {\it et al.},
``TESLA Technical Design Report Part III: Physics at an $\mathrm{e^+e^-}$
Linear Collider,''
hep-ph/0106315.

\bibitem{Martin:2000ww}
A.~D.~Martin, R.~G.~Roberts, W.~J.~Stirling and R.~S.~Thorne,
Eur.\ Phys.\ J.\ C {\bf 14} (2000) 133
[hep-ph/9907231].

\bibitem{Dittmar:1997md}
M.~Dittmar, F.~Pauss and D.~Z\"urcher,
Phys.\ Rev.\ D {\bf 56} (1997) 7284
[hep-ex/9705004].

\bibitem{Berends:1985qa}
F.~A.~Berends and R.~Kleiss,
Z.\ Phys.\ C {\bf 27} (1985) 365.

\bibitem{Baur:1999kt}
U.~Baur, S.~Keller and D.~Wackeroth,
Phys.\ Rev.\ D {\bf 59} (1999) 013002
[hep-ph/9807417].

\bibitem{Wackeroth:1997hz}
D.~Wackeroth and W.~Hollik,
Phys.\ Rev.\ D {\bf 55} (1997) 6788
[hep-ph/9606398].

\bibitem{Hamberg:1991np}
R.~Hamberg, W.~L.~van Neerven and T.~Matsuura,
Nucl.\ Phys.\ B {\bf 359} (1991) 343;\\
W.~L.~van Neerven and E.~B.~Zijlstra,
Nucl.\ Phys.\ B {\bf 382} (1992) 11.

\bibitem{Baur:2000hm}
U.~Baur and T.~Stelzer,
Phys.\ Rev.\ D {\bf 61} (2000) 073007
[hep-ph/9910206].

\bibitem{Ellis:1997sc}
R.~K.~Ellis, D.~A.~Ross and S.~Veseli,
Nucl.\ Phys.\ B {\bf 503} (1997) 309
[hep-ph/9704239].

\bibitem{Catani:2000zg}
S.~Catani {\it et al.},
in {\sl Proceedings of the workshop on standard model physics (and more)
at the LHC}, eds.\ G.~Altarelli and M.~L.~Mangano,
(CERN 2000-04, Geneva, 2000), p.~1
[hep-ph/0005114].

\bibitem{Haywood:1999qg}
S.~Haywood {\it et al.},
in {\sl Proceedings of the workshop on standard model physics (and more)
at the LHC}, eds.\ G.~Altarelli and M.~L.~Mangano,
(CERN 2000-04, Geneva, 2000), p.~117
[hep-ph/0003275].

\bibitem{Denner:1993kt}
A.~Denner,
Fortsch.\ Phys.\  {\bf 41} (1993) 307.

\bibitem{Bardin:1988xt}
D.~Y.~Bardin, A.~Leike, T.~Riemann and M.~Sachwitz,
Phys.\ Lett.\ B {\bf 206} (1988) 539.

\bibitem{Beenakker:1997kn}
W.~Beenakker {\it et al.},
Nucl.\ Phys.\ B {\bf 500} (1997) 255
[hep-ph/9612260].

\bibitem{Burkhardt:1995tt}
H.~Burkhardt and B.~Pietrzyk,
Phys.\ Lett.\ B {\bf 356} (1995) 398;\\
S.~Eidelman and F.~Jegerlehner,
Z.\ Phys.\ C {\bf 67} (1995) 585
[hep-ph/9502298].

\bibitem{Kublbeck:1990xc}
J.~K\"ublbeck, M.~B\"ohm and A.~Denner,
Comput.\ Phys.\ Commun.\  {\bf 60} (1990) 165;\\
H.~Eck and J.~K\"ublbeck, {\it Guide to FeynArts 1.0\/}, 
University of W\"urzburg, 1992.

\bibitem{Passarino:1979jh}
G.~Passarino and M.~Veltman,
Nucl.\ Phys.\ B {\bf 160} (1979) 151.

\bibitem{Mertig:1991an}
R.~Mertig, M.~B\"ohm and A.~Denner,
Comput.\ Phys.\ Commun.\  {\bf 64} (1991) 345;\\
R.~Mertig, {\it Guide to FeynCalc 1.0\/}, University of W\"urzburg, 1992.

\bibitem{'tHooft:1979xw}
G.~'t Hooft and M.~Veltman,
Nucl.\ Phys.\ B {\bf 153} (1979) 365;\\
W.~Beenakker and A.~Denner,
Nucl.\ Phys.\ B {\bf 338} (1990) 349;\\
A.~Denner, U.~Nierste and R.~Scharf,
Nucl.\ Phys.\ B {\bf 367} (1991) 637.

\bibitem{Denner:1995xt}
A.~Denner, S.~Dittmaier and G.~Weiglein,
Nucl.\ Phys.\ B {\bf 440} (1995) 95
[hep-ph/9410338].

\bibitem{Passera:1998uj}
M.~Passera and A.~Sirlin,
Phys.\ Rev.\ D {\bf 58} (1998) 113010
[hep-ph/9804309].

\bibitem{Sirlin:1980nh}
A.~Sirlin,
Phys.\ Rev.\ D {\bf 22} (1980) 971;\\
W.~J.~Marciano and A.~Sirlin,
Phys.\ Rev.\ D {\bf 22} (1980) 2695
[Erratum-ibid.\ D {\bf 31} (1980) 213] and
Nucl.\ Phys.\ B {\bf 189} (1981) 442.

\bibitem{Stuart:1991xk}
R.~G.~Stuart,
Phys.\ Lett.\ B {\bf 262} (1991) 113;\\
A.~Aeppli, F.~Cuypers and G.~J.~van Oldenborgh,
Phys.\ Lett.\ B {\bf 314} (1993) 413
[hep-ph/9303236];\\
H.~Veltman,
Z.\ Phys.\ C {\bf 62} (1994) 35.

\bibitem{Denner:2000bj}
A.~Denner, S.~Dittmaier, M.~Roth and D.~Wackeroth,
Nucl.\ Phys.\ B {\bf 587} (2000) 67
[hep-ph/0006307].

\bibitem{Jadach:1998hi}
S.~Jadach, W.~Placzek, M.~Skrzypek, B.~F.~Ward and Z.~Was,
Phys.\ Lett.\ B {\bf 417} (1998) 326
[hep-ph/9705429]; \\
W.~Beenakker, F.~A.~Berends and A.~P.~Chapovsky,
Nucl.\ Phys.\ B {\bf 548} (1999) 3
[hep-ph/9811481]; \\
Y.~Kurihara, M.~Kuroda and D.~Schildknecht,
Nucl.\ Phys.\ B {\bf 565} (2000) 49
[hep-ph/9908486].

\bibitem{Grunewald:2000ju}
M.~W.~Gr\"unewald {\it et al.},
in {\it Reports of the Working Groups on Precision Calculations
for LEP2 Physics}, eds.\ S.~Jadach, G.~Passarino and R.~Pittau
(CERN 2000-009, Geneva, 2000), p.~1
[hep-ph/0005309].

\bibitem{Melnikov:1996fx}
K.~Melnikov and O.~Yakovlev,
Nucl.\ Phys.\ B {\bf 471} (1996) 90
[hep-ph/9501358];\\
W.~Beenakker, F.~A.~Berends and A.~P.~Chapovsky,
Nucl.\ Phys.\ B {\bf 508} (1997) 17
[hep-ph/9707326].

\bibitem{Denner:1998ia}
A.~Denner, S.~Dittmaier and M.~Roth,
Nucl.\ Phys.\ B {\bf 519} (1998) 39
[hep-ph/9710521].


\bibitem{Argyres:1995ym}
E.~N.~Argyres {\it et al.},
Phys.\ Lett.\ B {\bf 358} (1995) 339
[hep-ph/9507216].

\bibitem{Baur:1995aa}
U.~Baur and D.~Zeppenfeld,
Phys.\ Rev.\ Lett.\  {\bf 75} (1995) 1002
[hep-ph/9503344].

\bibitem{Dittmaier:1999nn}
S.~Dittmaier,
Phys.\ Rev.\ D {\bf 59} (1999) 016007
[hep-ph/9805445].

\bibitem{Berends:1982uq}
F.~A.~Berends, R.~Kleiss, P.~De Causmaecker, R.~Gastmans, W.~Troost and T.~T.~Wu,
Nucl.\ Phys.\ B {\bf 206} (1982) 61;\\
R.~Kleiss,
Z.\ Phys.\ C {\bf 33} (1987) 433.

\bibitem{Dittmaier:2000mb}
S.~Dittmaier,
Nucl.\ Phys.\ B {\bf 565} (2000) 69
[hep-ph/9904440].

\bibitem{Catani:1996jh}
S.~Catani and M.~H.~Seymour,
Phys.\ Lett.\ B {\bf 378} (1996) 287
[hep-ph/9602277] and
Nucl.\ Phys.\ B {\bf 485} (1997) 291
[Erratum-ibid.\ B {\bf 510} (1997) 291]
[hep-ph/9605323].

\bibitem{Kripfganz:1988bd}
J.~Kripfganz and H.~Perlt,
Z.\ Phys.\ C {\bf 41} (1988) 319;\\
H.~Spiesberger,
Phys.\ Rev.\ D {\bf 52} (1995) 4936
[hep-ph/9412286].

\bibitem{Caso:1998tx}
C.~Caso {\it et al.}  [Particle Data Group Collaboration],
Eur.\ Phys.\ J.\ C {\bf 3} (1998) 1.

\bibitem{Lai:1997mg}
H.~L.~Lai {\it et al.},
Phys.\ Rev.\ D {\bf 55} (1997) 1280
[hep-ph/9606399].

\bibitem{KLN}
T.~Kinoshita,
J.\ Math.\ Phys.\ {\bf 3} (1962) 650; \\
T.~D.~Lee and M.~Nauenberg,
Phys.\ Rev.\ {\bf 133} (1964) B1549.

\bibitem{Zykunov:2001mn}
V.~A.~Zykunov,
hep-ph/0107059.

\end{thebibliography}
\end{document}